\documentclass[5p,twocolumn,number,sort&compress,times]{elsarticle}

\journal{Nuclear Instruments and Methods in Physics Research A}

\usepackage{datetime}
\usepackage{lineno}

\usepackage{graphics}
\usepackage{color}
\usepackage{graphicx}
\usepackage{amssymb}
\usepackage{amsmath}
\usepackage{epstopdf}
\usepackage{xspace}

\usepackage{mathbbol}
\usepackage{wasysym} 
\usepackage{bbm}     
 
\usepackage{rotating}
\usepackage{dcolumn}% Align table columns on decimal point
\usepackage{bm}% bold math
\usepackage{longtable}% Include figure files
\usepackage{multirow}
\usepackage{enumitem}

\usepackage{ifpdf}
\ifpdf
\usepackage[pdftex]{hyperref}
\else
\usepackage[hypertex]{hyperref}
\fi

\hypersetup{
 pdftitle={},%
 pdfauthor={},%
 pdfsubject={},%
 pdfkeywords={},%
 pdfstartview={},%
 bookmarksopen=true, breaklinks=true, debug=true,
 colorlinks=true, linkcolor=blue, citecolor=red, urlcolor=red,
 hyperfigures=true
}

\def\figwid{3.50in}

\def\FThing#1#2{#1~\ref{#2}}

\def\Thing#1#2{#1.~\ref{#2}}
\def\Things#1#2{\Thing{#1s}{#2}}

\def\Sec#1{\Thing{Sect}{#1}}

\def\Fig#1{\Thing{Fig}{#1}}
\def\Figs#1{\Things{Fig}{#1}}

\def\Figure#1{\FThing{Figure}{#1}}

\def\Tab#1{\FThing{Table}{#1}}

\def\Eq#1{Eq.~(\ref{#1})}

\def\beq{\begin{linenomath}\begin{equation}}
\def\eeq{\end{equation}\end{linenomath}}
\def\beqa{\beq\begin{aligned}}
\def\eeqa{\end{aligned}\eeq}

\newcommand{\mi}  {~\ensuremath{{\mu\mathrm{m}}     }}

\newcommand{\mm}  {~\ensuremath{{\mathrm{mm}}     }}

\newcommand{\ie}   {{\it i.e.,}~}

\newcommand{\etal} {{\it {et al.}}}

\newcommand{\kev}  {~\ensuremath{{\mathrm{keV}}     }}
\newcommand{\gev}  {~\ensuremath{{\mathrm{GeV}}     }}

\newcommand{\refone}{\cite{ref:xbsmnim}}
\newcommand{\reftwo}{\cite{ref:xbsmnimtwo}}
\newcommand{\refboth}{\cite{ref:xbsmnim,ref:xbsmnimtwo}}
\newcommand{\Refone}{Ref.~\refone}

\newcommand{\degree}{$^\circ$}

\newcommand{\amp}{\ensuremath{A_b}}
\newcommand{\ampta}{\ensuremath{{\left<{A_b}\right>}}}
\newcommand{\siz}{\ensuremath{{\sigma_b}}}
\newcommand{\sizta}{\ensuremath{{\left<{\sigma_b}\right>}}}

\newcommand{\off}{\ensuremath{{y_b}}}
\newcommand{\Eb}{\ensuremath{{E_b}}}

\newcommand{\ap}{\ensuremath{{a^{\,\prime}}}}

\newcommand{\lp}{\ensuremath{{L^{\,\prime}}}}

\def\prftt{{\tt prf}}

\begin{document}

\title{Operation of the CESR-TA vertical beam size monitor at \Eb=4~GeV}

%\setpagewiselinenumbers
%%%\modulolinenumbers[1]
%\linenumbers
%\switchlinenumbers

\renewcommand{\thefootnote}{\fnsymbol{footnote}}

\author[cor]{J.~P.~Alexander}
\author[cor]{C.~Conolly}
\author[cor]{E.~Edwards}
\author[kek,tsu]{J.~W.~Flanagan}
\author[cor]{E.~Fontes}
\author[cor]{B.~K.~Heltsley\corref{cor1}}\ead{bkh2@cornell.edu}
\author[cor]{A.~Lyndaker}
\author[cor]{D.~P.~Peterson}
\author[cor]{N.~T.~Rider}
\author[cor]{D.~L.~Rubin}
\author[cor]{R.~Seeley}
\author[cor]{J.~Shanks}
\address[cor]{Cornell University, Ithaca, NY 14853, USA}
\address[kek]{High Energy Accelerator Research Organization (KEK),
  Tsukuba, Japan}
\address[tsu]{Department of Accelerator Science, Graduate University for 
              Advanced Studies (SOKENDAI), Tsukuba, Japan}
\cortext[cor1]{Corresponding author}

\begin{abstract}
{\small
We describe operation of the CESR-TA vertical beam size monitor (xBSM) 
with $e^\pm$ beams with \Eb=4~GeV. The xBSM measures vertical beam size by 
imaging synchrotron radiation x-rays through an optical element
onto a detector array of 32 InGaAs photodiodes with 50$~\mu$m pitch.
The device has previously been successfully used to measure vertical 
beam sizes of $10-100~\mu$m on a bunch-by-bunch, turn-by-turn basis at 
$e^\pm$ beam energies of $\sim$2~GeV and source magnetic fields below 2.8~kG, 
for which the detector required calibration for incident x-rays of 1-5\kev.
At $E_{\rm b}=4.0~$GeV and $B$=4.5~kG, 
however, the incident synchrotron radiation spectrum extends to $\sim$20~keV, 
requiring calibration of detector response in that regime. Such a calibration
is described and then used to analyze data taken with several different
thicknesses of filters in front of the detector. We obtain a relative 
precision of better than 4\% on beam size measurement from 15-100~$\mu$m 
over several different ranges of x-ray energy, including both
1-12~keV and 6-17~keV. The response of an identical detector, but tilted
vertically by 60\degree\ in order to increase magnfication without a longer
beamline, is measured and shown to improve x-ray detection above 4\kev\ without
compromising sensitivity to beam size. 
We also investigate operation of a coded aperture
using gold masking backed by synthetic diamond.
}
\end{abstract}

\date{\today}
\maketitle

\renewcommand{\thefootnote}{\arabic{footnote}}

\section[Introduction]{Introduction}

\label{sec:intro}

\begin{figure}[t]
   \includegraphics*[width=\figwid]{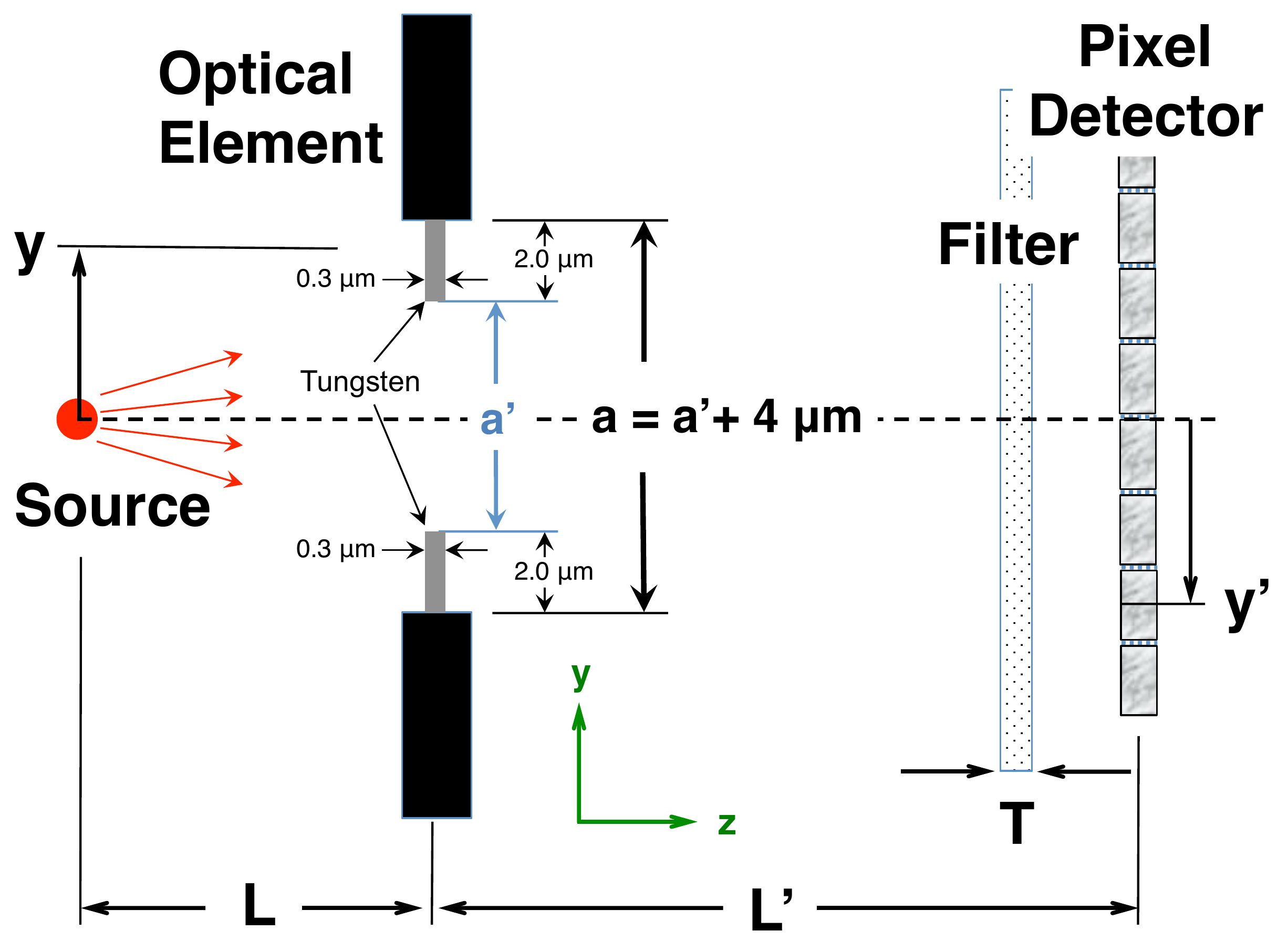}
   \caption{Simplified schematic of xBSM layout (not to scale), shown with a
     model of the adjustable pinhole (PH) optical element. Dimensions appear in
     \Tab{tab:geom}.}
\label{fig:geom}
\end{figure}

The CESR-TA x-ray beam size monitor~\cite{ref:xbsmnim,ref:xbsmnimtwo}
(xBSM) images synchrotron radiation from a hard-bend magnet through a 
single- or multi-slit optical element onto a 32-strip photodiode 
detector~\cite{ref:fermionics}
with 50\mi\ pitch and sub-ns response. A simplified schematic of the 
CESR-TA xBSM setup is shown in \Fig{fig:geom}, and dimensions appear in 
\Tab{tab:geom}. Separate installations exist for electrons and positrons.
The hard-bend magnets have a bending radius of 31.1~m. The vertical beam size
\siz\ is extracted from a fit of the image for each bunch and turn to
a set of unit-area templates; each template has convolved the point response 
function (\prftt) with Gaussian smearing for a particular beam size and 
magnification. Each such fit also yields a vertical beam position 
\off\ and amplitude \amp. \Refone\ describes our use of both single-slit 
(pinhole) and multi-slit optical elements, the latter of which are known 
as coded apertures, for $e^\pm$ beam energies of \Eb=1.8-2.6\gev. 
In order to accurately predict the point response functions, in 
\refone\ we determined the detector's spectral response to 
x-rays up to about 5\kev. In \reftwo, we gave a detailed description 
of the design and operation of two silicon-backed coded apertures.

\begin{table}[t]
\small\selectfont
\caption{Geometrical parameters defining the CESR-TA xBSM 
         beamlines. Geometrical quantities are defined in \Fig{fig:geom}.
         Distances assume the coded aperture optic; the pinhole optic is 
         25\mm\ closer to the source point and hence has a magnification 
         value about 1\% larger than shown. The uncertainties on $L$ are 
         from an optical survey. The uncertainties on $\lp$ are from
         the survey, CESR orbit, and the associated depth of field. }
\label{tab:geom}
\setlength{\tabcolsep}{1.00pc}
\begin{center}
\small\selectfont
\begin{tabular}{ccc}
\hline\hline
\rule[10pt]{-1mm}{0mm}
 Parameter & $e^-$ beamline & $e^+$ beamline \\
\hline
\rule[10pt]{-1mm}{0mm}
 $L$ & $4356.5\pm3.9$\mm & $4485.2\pm4.0$\mm \\
 $\lp$ & $10621.1\pm1.0$\mm& $10011.7\pm1.0$\mm\\
 $M\equiv\lp/L$ & $2.4380\pm0.0022$ &  $2.2322\pm0.0020$\\
 \ap & $\approx 50-300$\mi & same as $e^-$\\
 $a$ & $\approx 50-1000$\mi & same as $e^-$\\
 $2\theta_{\rm max}=\ap/L$ & $11-69~\mu$rad &  $11-67~\mu$rad \\
\hline\hline
\end{tabular}
\end{center}
\end{table}

\begin{figure}[b]
   \includegraphics*[width=\figwid]{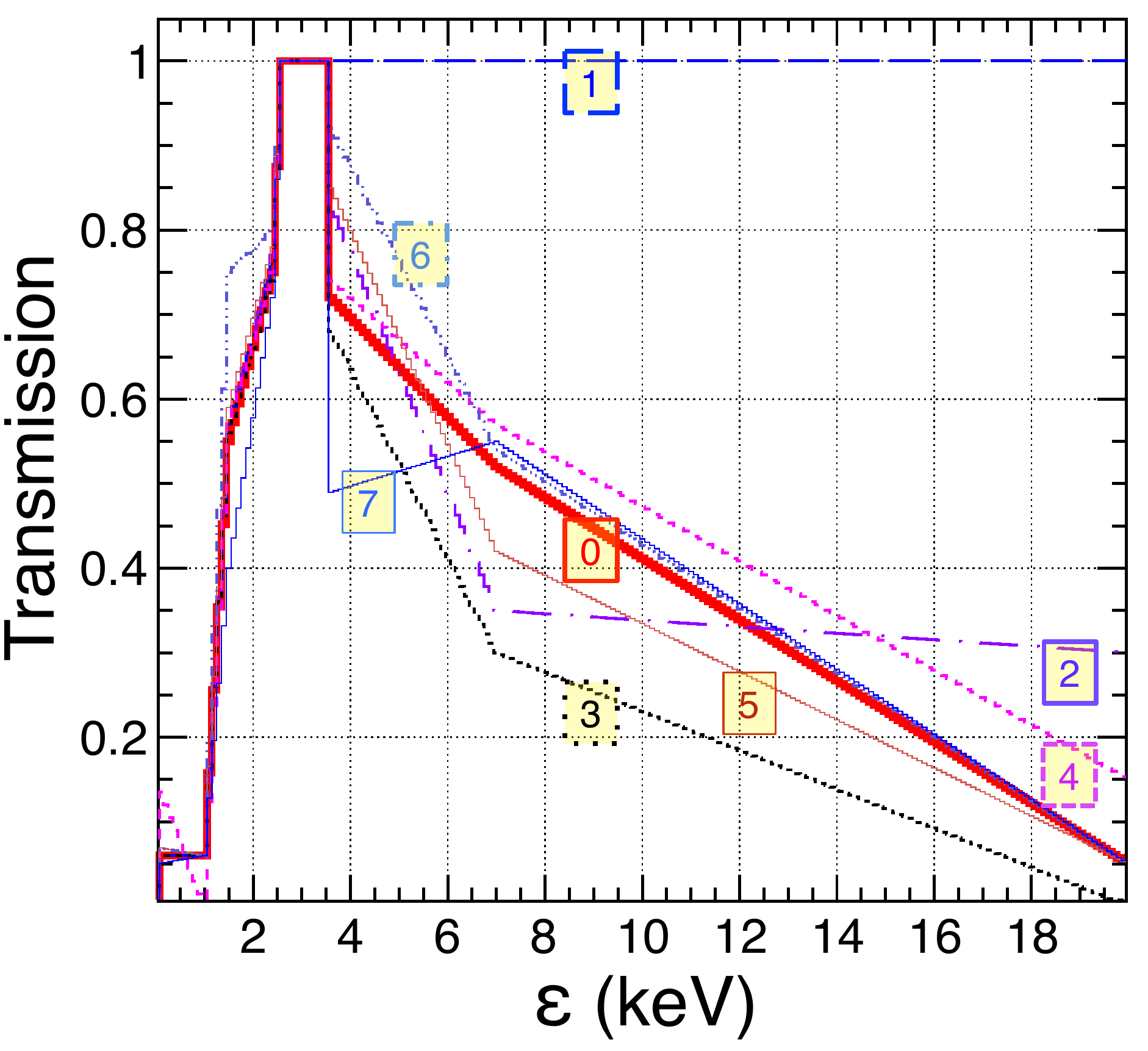}
   \caption{Spectral response curves used to explore sensitivity of measured
     beam size to the detected x-ray spectrum. The thickest solid curve
     represents the nominal spectrum and is labeled as ``0''; seven other
     alternative responses consistent with the data of 
     \refone\ are also shown and labeled with numbers from 1-7.}
\label{fig:specone}
\end{figure}

\begin{figure}[t]
   \includegraphics*[width=\figwid]{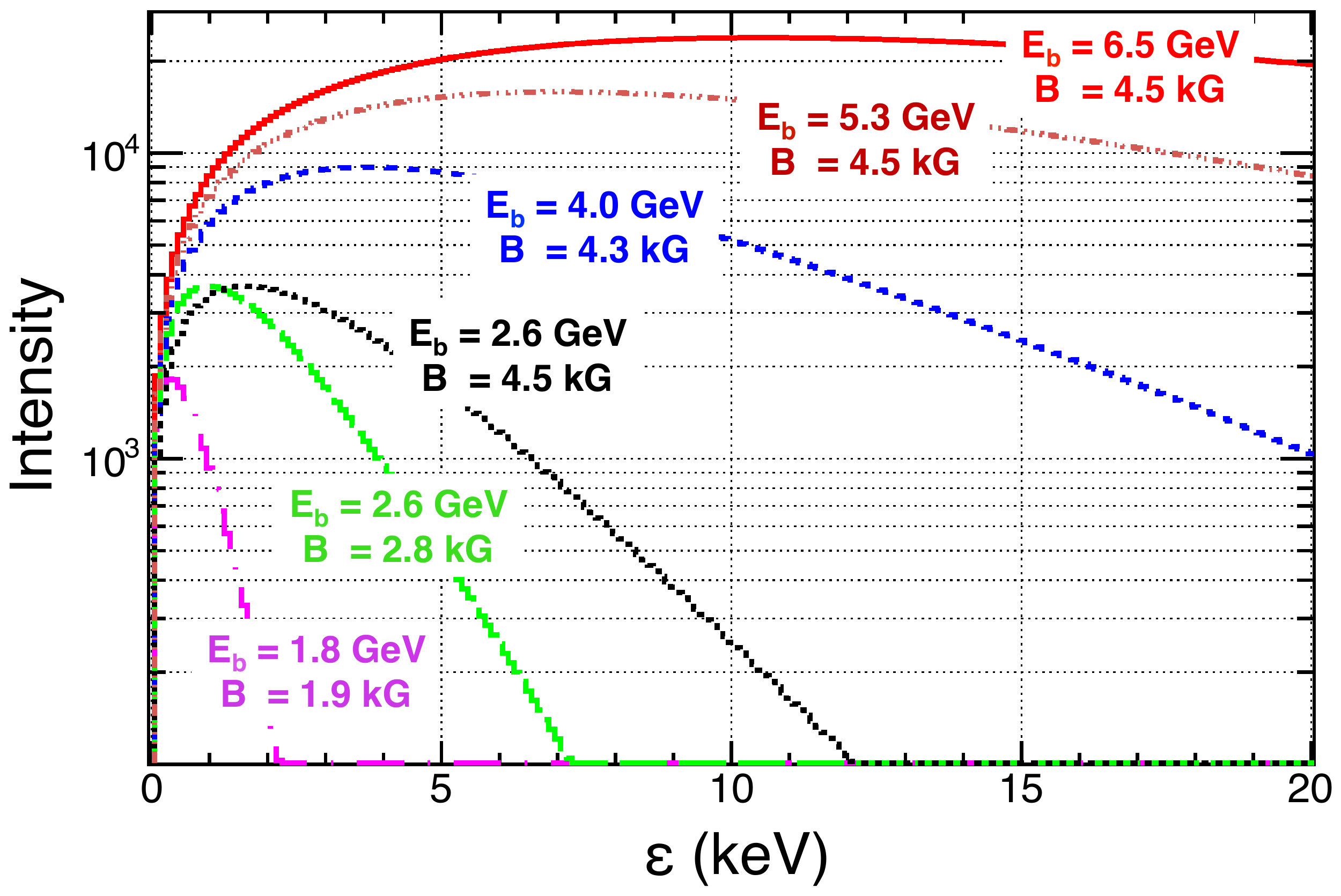}
   \caption{Spectra of x-rays for several $e^\pm$ beam energy/magnetic field 
     combinations, including a relative factor of $\rho/\gamma$ (bend
     radius/$(\Eb/m_e)$) to account for source length due to the $1/\gamma$
     horizontal width of the synchrotron radiation searchlight. The curves
     labeled with magnetic fields below 4.5~kG correspond to the current xBSM
     configuration; the 4.5~kG curves correspond to a planned upgrade with a
     dedicated, fixed-field source magnet.}
\label{fig:spectrum}
\end{figure}

Extending characterization of our detector's spectral response above 5\kev\ is
useful for two reasons. First, a similar detector is an option for beam 
size measurement using x-rays from the SuperKEKB positron (\Eb=4\gev) and 
electron (\Eb=7\gev) beams. 
Second, a new, dedicated xBSM installation at CESR is being 
designed. Unlike the current setup, it will be able to operate in both
CESR-TA (beam energies of $\sim$2\gev) and CHESS (\Eb=5.3-6.5\gev) operating 
conditions. In both scenarios a substantial fraction of the x-ray spectrum 
will be above 5\kev, as is evident in \Fig{fig:spectrum} when comparing
existing (B$<$4.5~kG) and planned (B=4.5~kG) installations. 
In \Fig{fig:specone}, the response as determined in 
\refone\ is shown; note that the data acquired for 
\Eb=1.8-2.6\gev\ could not distinguish between the possible response shown 
as curve \#1 from that of curve \#3. 

The balance of this paper is organized as follows. First, in \Sec{sec:gap},
the opening in the adjustable pinhole optical element for data acquired at
\Eb=4.0\gev\ is determined using techniques similar to those used in 
\refone\ for operation near \Eb=2\gev. Second, \Sec{sec:spec} describes 
measurement of the detector spectral response using \Eb=4.0\gev\ data,
again using techniques previously described~\refone. Third, in 
\Sec{sec:size} we describe beam size measurements using several different 
x-ray ranges and a large range of beam sizes. Fourth, \Sec{sec:tilt} reports on
operation and calibration of the detector after tilting it vertically by 
60\degree\ with respect to the incident x-rays. Finally, \Sec{sec:cahe} 
describes evaluation of a prototype for a diamond-backed 
coded aperture for use at SuperKEKB, and \Sec{sec:conc} summarizes our 
conclusions.

\begin{figure}[b]
   \includegraphics*[width=\figwid]{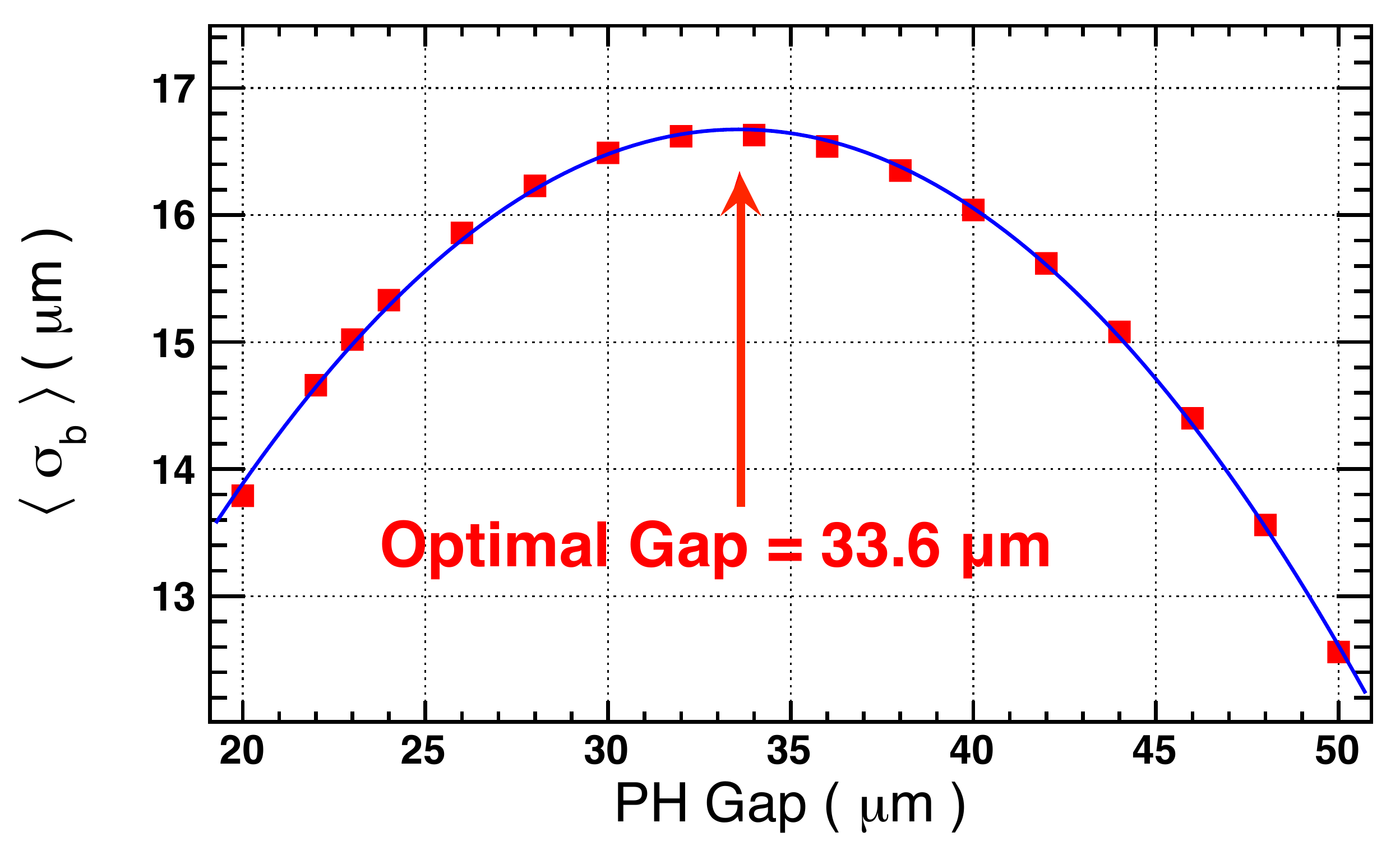}
   \caption{Turn-averaged beam size \sizta\ for the same run at \Eb=4.0\gev\ 
     based on \prftt's using the specified value of \ap\ with a parabolic fit 
     to the points superimposed.}
\label{fig:gapscan}
\end{figure}

\section[Size of pinhole opening]{Size of pinhole opening}
\label{sec:gap}

To determine the size of the gap in the adjustable pinhole (PH) used for
our filter studies at \Eb=4\gev, we follow a procedure similar to that 
described in 
Sect.~5.3 of Ref.~\cite{ref:xbsmnim}, and apply it to \Eb=4\gev\ data and the
corresponding predicted synchrotron radiation spectrum. We need to determine,
for the PH model shown in \Fig{fig:geom}, gap size \ap\ used for data acquired 
at \Eb=4\gev. We know the motor setting ($mset=4.22$), but need to determine 
its calibration to absolute distance.

The first step is to find the opening \ap\ our model predicts to result in 
the narrowest \prftt\ for an incident x-ray spectrum for magnet field 
strength B=0.43~T (which preserves the required bending radius at 
\Eb=4.0\gev). Each point in \Fig{fig:gapscan} corresponds to the beam 
size reconstructed from the same 1024-turn run using a \prftt\ made with a 
different value of \ap. The narrowest \prftt\ corresponds to the largest 
reconstructed beam size at \ap=33.6\mi. This becomes the reference value 
for amplitude.\footnote{For this analysis, a measured value of ``amplitude'' 
represent the factor, divided by the bunch current in mA independently 
measured for that run, multiplying a unit-area template so as to yield a 
best-fit to the image data points.}

\begin{figure}[t]
   \includegraphics*[width=\figwid]{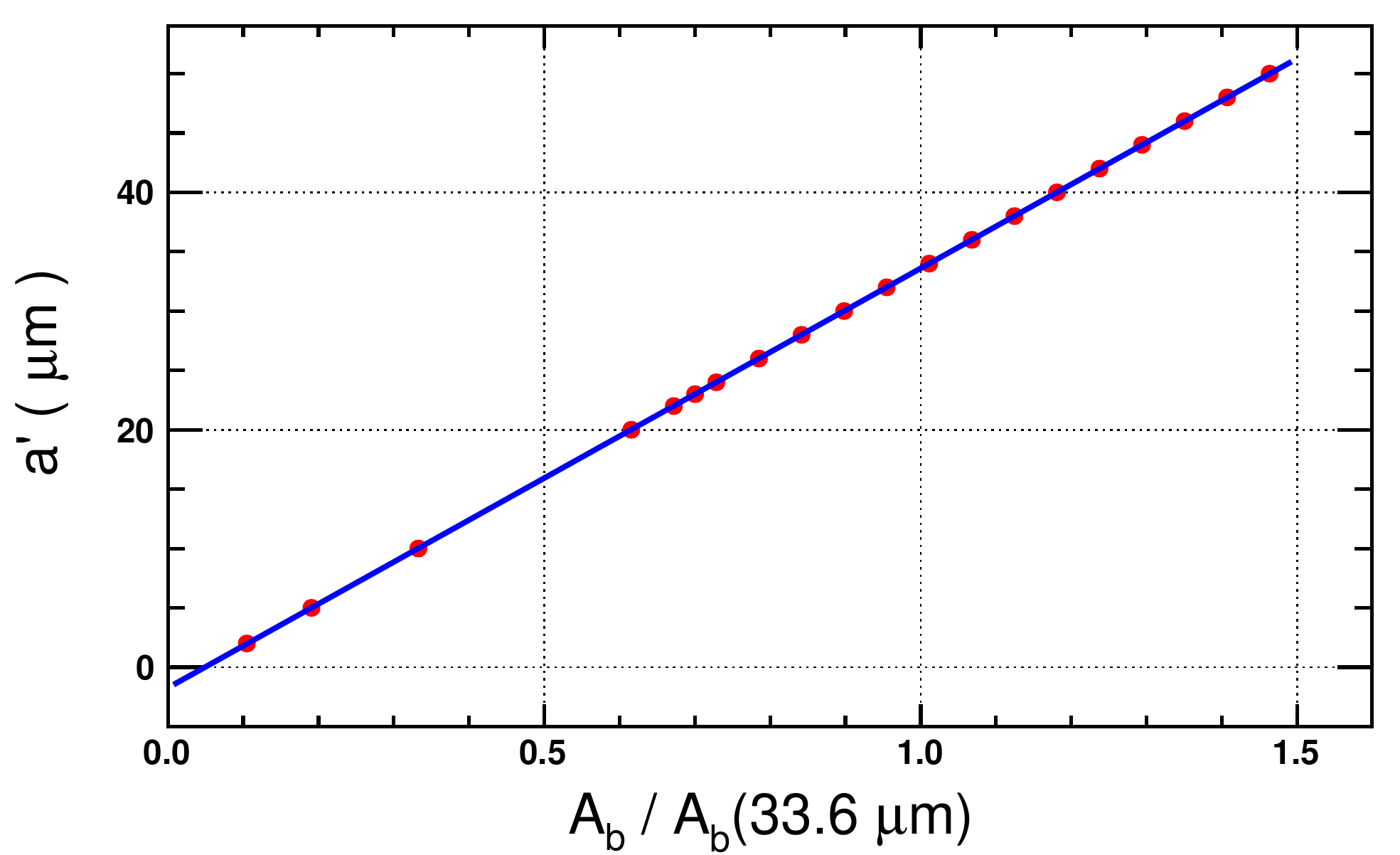}
   \caption{Model prediction for the pinhole gap opening \ap~vs.~image 
     amplitude \amp (relative to the value at \ap=33.6\mi), with a 
     linear fit to the points superimposed, all at \Eb=4.0\gev.}
\label{fig:Formula}
\end{figure}
\begin{figure}[t]
   \includegraphics*[width=\figwid]{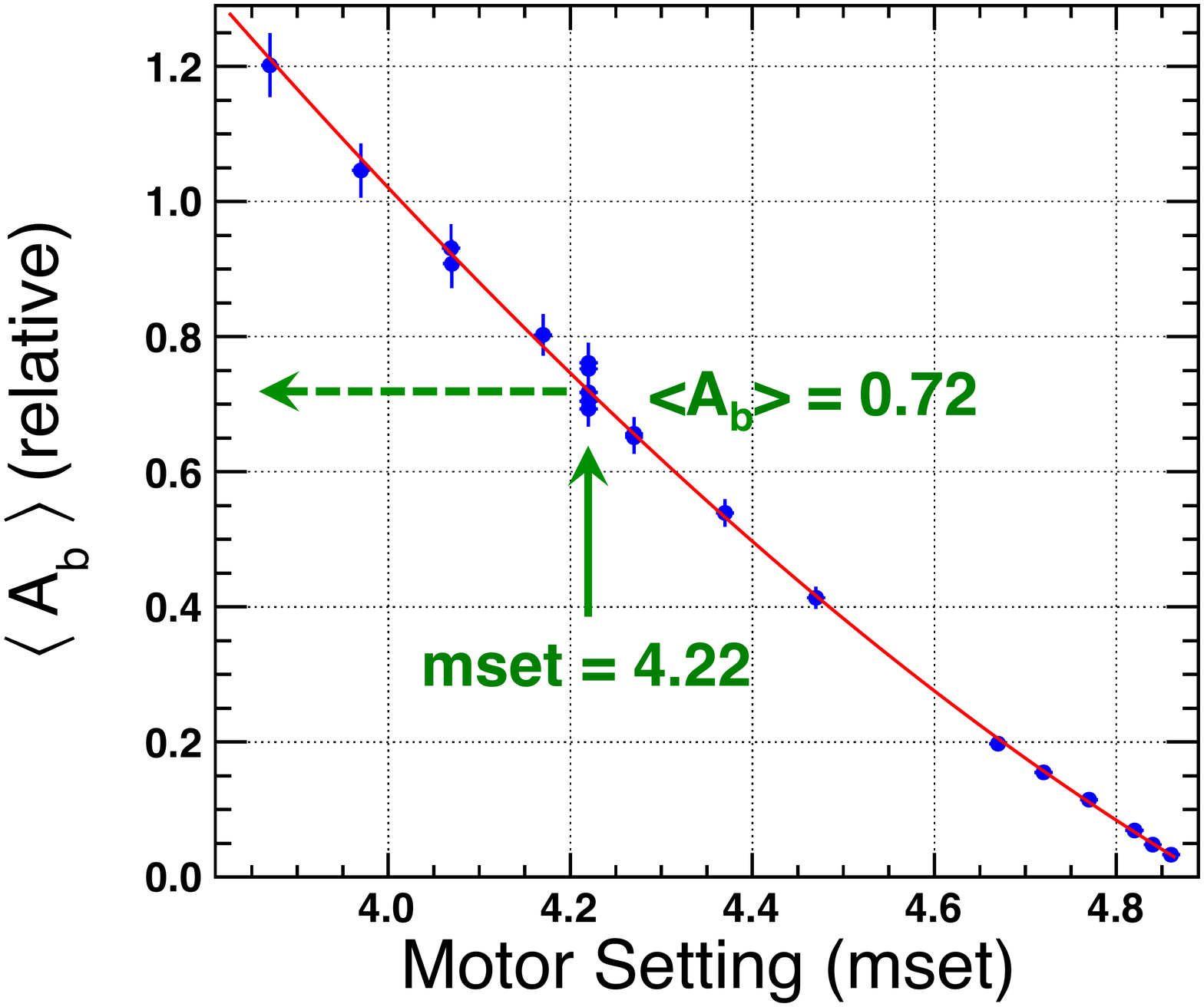}
   \caption{Relative turn-averaged image amplitude \ampta\ vs.~motor setting, 
     $mset$, from the xBSM dataset taken at \Eb=4.0\gev. The overlaid curve
     represents a second-order polynomial fit to the data points.}
\label{fig:tgcalfour}
\end{figure}

The second step is to determine the formula relating relative image amplitude
to the opening \ap. \Fig{fig:Formula} is generated from the expected x-ray
spectrum, assumed spectral response, and the PH model, resulting in\footnote{The
  formula in \Eq{eqn:form} has a non-zero offset due to the
  partially-transmitting ``lips'' at the two edges of the gap as shown in 
  \Fig{fig:geom}; \ie some light is transmitted even for \ap=0.} 
\beq
\ap(\mu{\rm m})=35.3\times \dfrac{\amp}{\amp(\ap=33.6\mi)} - 1.7.
\label{eqn:form}
\eeq

\begin{figure}[t]
   \includegraphics*[width=\figwid]{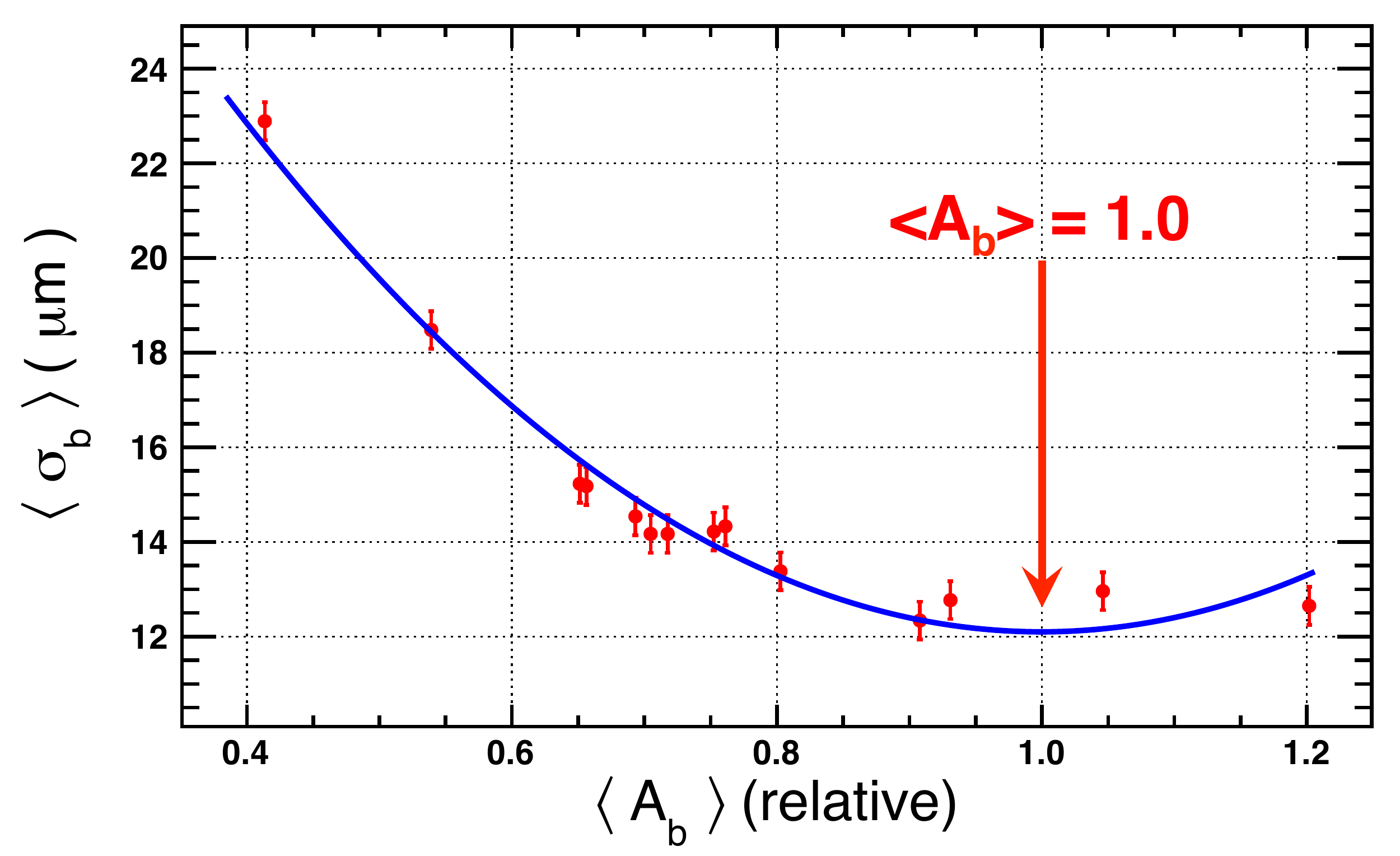}
   \caption{Turn-averaged beam size \sizta\ vs.~relative turn-averaged image 
     amplitude \ampta, from the xBSM dataset taken at \Eb=4.0\gev. The 
     overlaid curve represents a second-order polynomial fit to the data
     points.}
\label{fig:tasgapfour}
\end{figure}
\begin{figure}[b]
   \includegraphics*[width=\figwid]{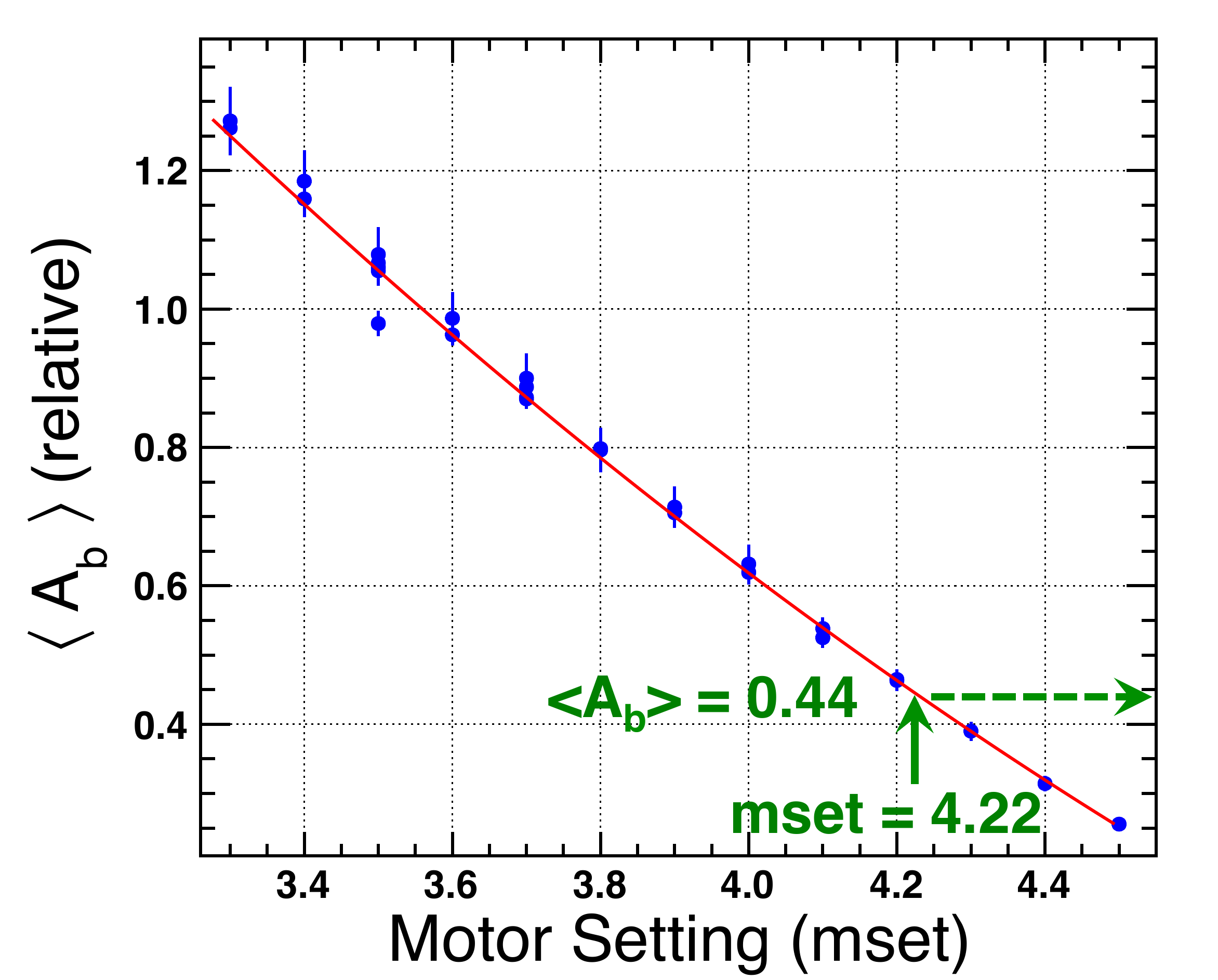}
   \caption{Relative turn-averaged image amplitude \ampta\ vs.~motor setting 
     $mset$ from the xBSM dataset taken at \Eb=2.1\gev. The overlaid curve
     represents a third-order polynomial fit to the data points.}
\label{fig:tgcaltwo}
\end{figure}

What remains is to determine the relative area of images for the actual opening
we used relative to the optimal value, \ap=33.6\mi. In order to do so, several
data runs were acquired with a variety of motor settings ($mset$) for the gap.
The motor setting is known to be not quite linear with PH opening; this is
apparent in \Fig{fig:tgcalfour} where the image amplitude is plotted
vs.~$mset$. The normalization for image amplitude comes from this same 
dataset, which can be seen in \Fig{fig:tasgapfour} to attain the narrowest 
image (by definition) at relative image size of 1.0. Using \Eq{eqn:form}, the
pinhole size, as extracted from \Eb=4.0\gev\ data, is $\ap=0.72\times
35.3-1.7=23.7$\mi, with an estimated uncertainty of 1.0\mi.

\begin{figure}[t]
   \includegraphics*[width=\figwid]{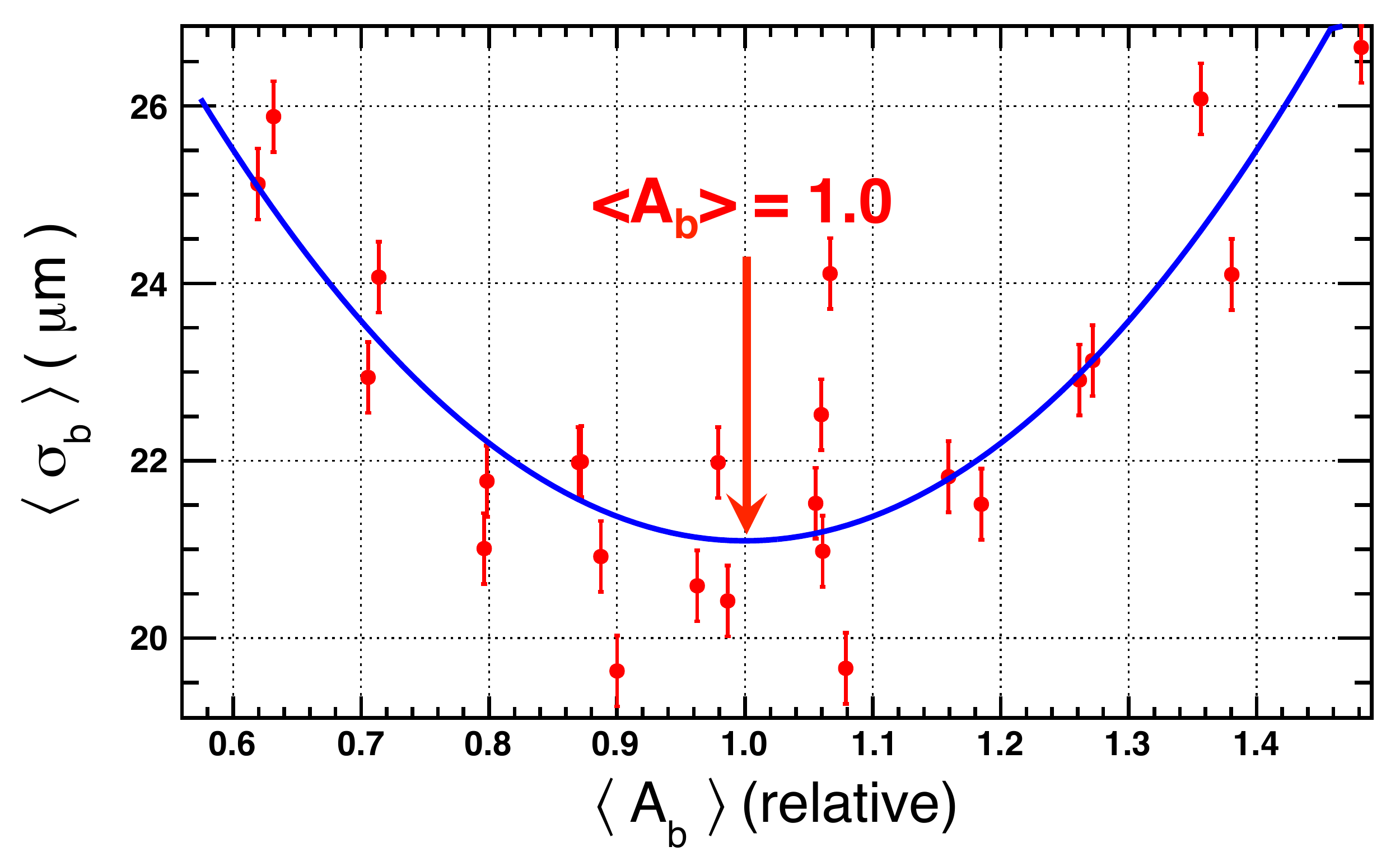}
   \caption{Turn-averaged beam size \sizta\ vs.~relative turn-averaged image 
     amplitude \ampta, from the xBSM dataset taken at \Eb=2.1\gev.}
\label{fig:tasgaptwo}
\end{figure}

An independent dataset was acquired at \Eb=2.1\gev\ for a similar purpose. At
this energy, the optimal \ap=50\mi~\cite{ref:xbsmnim} and Eq.~(19) from 
Ref.~\cite{ref:xbsmnim} applies. \Figs{fig:tgcaltwo} and \ref{fig:tasgaptwo} 
show the result of that $mset$ scan, which imply 
$\ap=0.44\times 50.6-0.6=21.7$\mi, 
with an estimated uncertainty of 0.5\mi\ (due to more data points acquired 
for large \amp). We weight the \Eb=2.1 and 4.0\gev\ results by their 
inverse square uncertainties to obtain \ap=22.1$\pm$0.5\mi. Further analysis of
this dataset uses a model with \ap=22\mi.\footnote{The value of \ap\ extracted 
from data requires an assumed detector spectral response with this
method because diffraction effects vary with both slit size and wavelength. 
But the detector spectral response is not known $a~priori$; it is determined 
from filter studies in \Sec{sec:spec}. Hence, the entire 
analysis of this and subsequent sections was repeated after starting with 
the response labeled \#3 in \Fig{fig:specone}; the result shown here is 
from the second iteration, which is 1\mi\ smaller than that of the first 
iteration.}

\section[Detector spectral response]{Detector spectral response}
\label{sec:spec}

We define the spectral response of the detector as the (relative) fraction 
of intensity recorded as pulse height as a function of wavelength. It includes
the effect of inactive material in front of the active pixels and the partial 
absorption as a function of wavelength of the pixels themselves. We measure it 
using data acquired with 
various filters in place, wherein each filter masks out some portion of the 
spectrum different from the other filters. The overall detector spectral 
response is part of a model described in \refone; the model 
is adjusted until the predicted integrated intensity on the detector matches 
the model. We take as our measure of integrated intensity the turn-averaged
fitted amplitude per unit beam current, \ampta, for a particular run. 
The turn-by-turn amplitude per unit beam current \amp\ is the coefficient 
multiplying a particular best fit, unit-area template that is also fitted 
for three other free parameters: beam size, beam position, and background
constant across the detector. This choice of \ampta\ is superior to the 
integrated pulse height because a correct response is still obtained for 
cases where the beam size is large and the image overflows the detector.
%(\ie ).

\begin{figure}[t]
   \includegraphics*[width=\figwid]{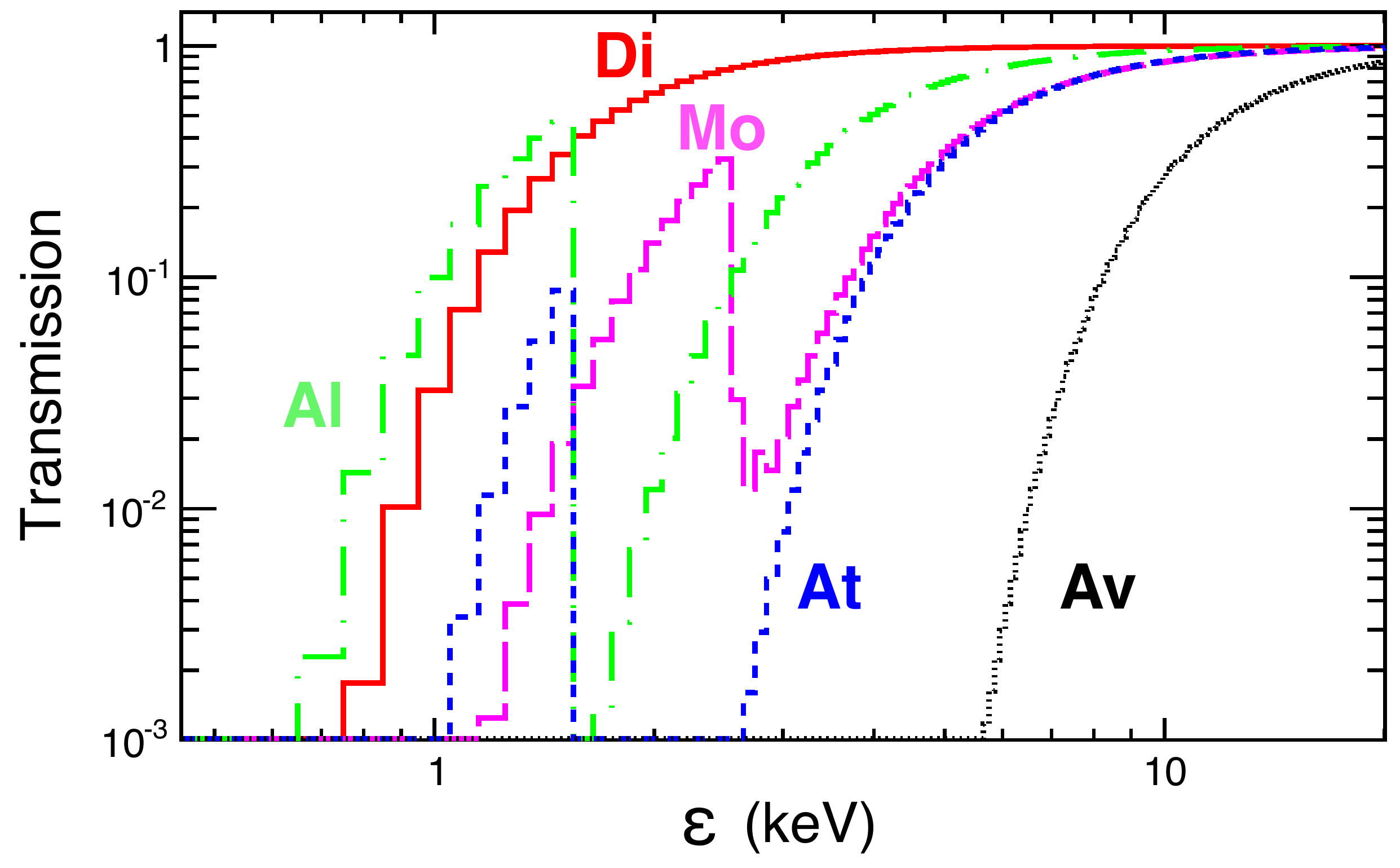}
   \caption{Transmission as a function of x-ray energy $\epsilon$ 
     for each of the filters used.}
\label{fig:trans}
\end{figure}

At each of two beam energies, \Eb=2.1\gev\ and 4.0\gev, data were acquired 
with no filter in place (NF) and with four or five different filters, shown in
\Tab{tab:filters}. The thickest filter, Av, was not used at \Eb=2.1\gev\ 
because it absorbs the entire incident spectrum. \Fig{fig:trans} and 
\Tab{tab:filters} together show the properties of each filter. In particular,
\Tab{tab:filters} includes the uncertainty in thickness and its
effect upon the uncertainty in energy deposition. This uncertainty was ignored
in \refone, where the resulting uncertainty in spectral response was therefore
underestimated. Here we account for the effect of that thickness uncertainty 
in the detector energy deposition from our model, as well as for several 
effects on the measured values of \ampta: uncertainty in current determination 
and precision, inherent run-to-run variations in the same CESR fill, and
fill-to-fill variations due to different horizontal illuminations~\refone.

\begin{table}[t]
\small\selectfont
\caption{Properties of each filter: its name, element, 
  atomic number ($A$), density ($\rho$), thickness ($T$), relative 
  uncertainty in thickness ($\Delta T/T$), relative change in energy 
  deposition per 10\% change in thickness ($D_{10}^{\Eb}$), 
  for both \Eb=2.1 and 4.0\gev.}
%and relative uncertainty
%  in energy deposition corresponding to the actual uncertainty in
%  thickness, $D$.}
\label{tab:filters}
\setlength{\tabcolsep}{0.52pc}
\begin{center}
\small\selectfont
\begin{tabular}{lccccccc}
\hline\hline
\rule[10pt]{-1mm}{0mm}
 Name & Ele- & $A$ & $\rho$ & $T$ & $\Delta T / T$  & $D_{10}^{2.1}$
 & $D_{10}^{4.0}$ \\
 &ment & & (g/cc) & ($\mu$m) & (\%) &  (\%) &  (\%) \\
\hline
\rule[10pt]{-1mm}{0mm}
Di & C  & 12.01 & 3.515 & 4.4  & 10 & 4.3 & 0.83 \\%& 0.83 \\
Mo & Mo & 95.94 & 10.28 & 1.91 & 3.1& 17 &  5.9  \\%& 1.83 \\
Al & Al & 26.98 & 2.694 & 7.2  & 5.0& 12 & 3.1  \\%& 1.55 \\
At & Al & 26.98 & 2.694 & 22.9 & 1.4& 22 & 4.1  \\%& 1.03 \\
Av & Al & 26.98 & 2.694 & 194  & 0.7& -  & 12 \\%& 1.70 \\
\hline\hline
\end{tabular}
\end{center}
\end{table}

\begin{figure}[t]
   \includegraphics*[width=\figwid]{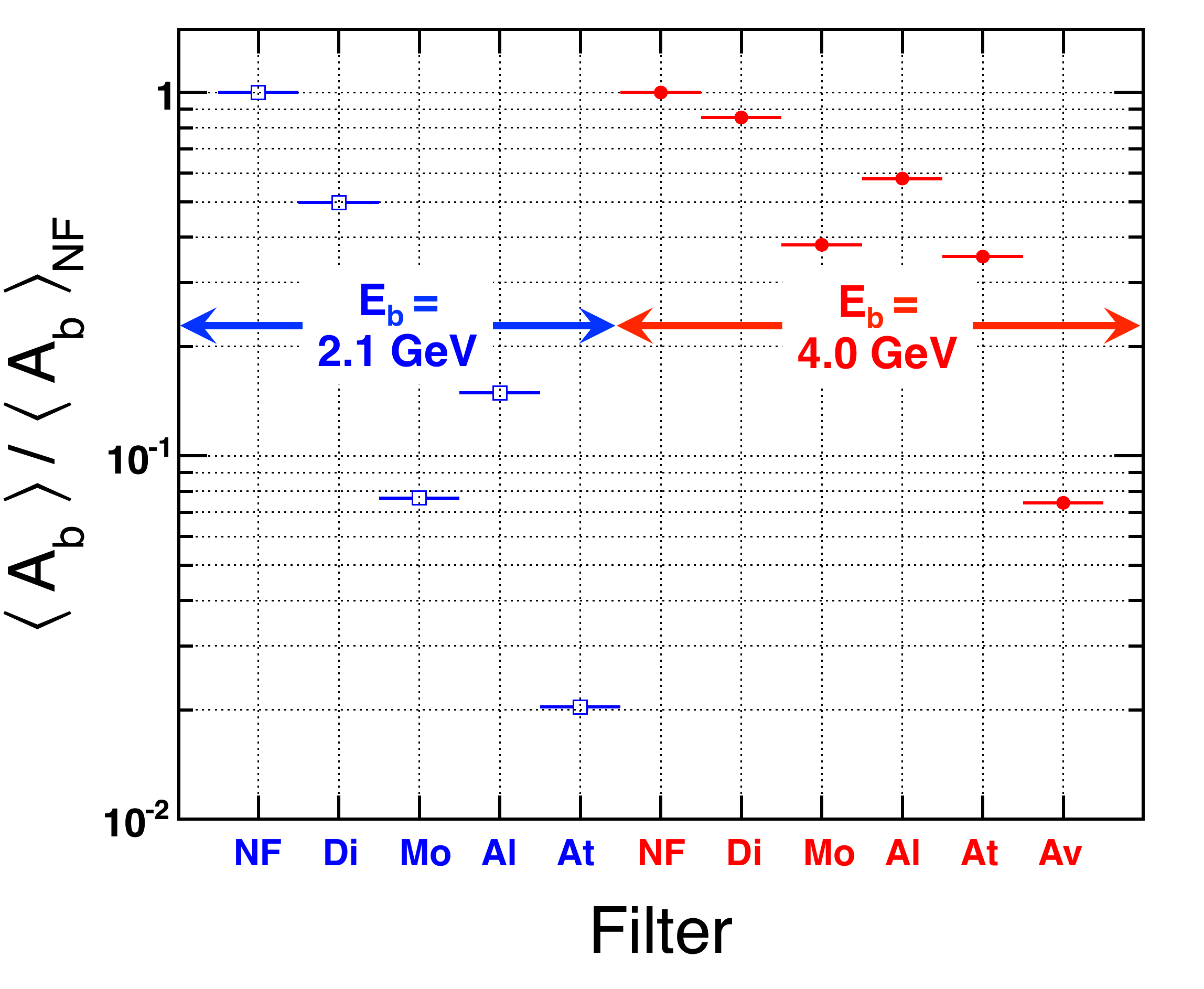}
   \caption{Measured turn-averaged amplitude \ampta\ using different filters,
     for \Eb=2.1\gev\ (left, open 
     squares) and 4.0\gev\ (right, closed circles). Each set has been 
     normalized to its own arbitrary scale where the no-filter data point 
     has a value of 1.0. Vertical error bars are smaller than the symbols.}
\label{fig:filterratios}
\end{figure}

The arbitrarily normalized turn-averaged amplitudes for both beam energies and
different filters in place are shown in \Fig{fig:filterratios}. Note that the
filters have much larger relative effects at lower beam energy where the energy
absorbed in the filter constitutes a larger fraction of a smaller amount of
incident energy. 

At this point it is convenient to define two image amplitude ratios, one for
data and one for the model. For a particular filter $f$ we define the ratio 
of the measured turn-averaged amplitude to the corresponding measured 
no-filter result as
\beq
R_{data}^f \equiv \frac{\ampta_f}{\ampta_{NF}}
\eeq
and the comparable ratio for the model as
\beq
R_{model}^f \equiv \frac{A_{b,f}^{model}}{A_{b,NF}^{model}}.
\eeq
We determine the spectral response of the detector by 
iterating the following process on our model:
\begin{itemize}
\item Adjust the model's spectral response, using a piecewise linear function.
\item Run the modified model to obtain the predicted energy deposited into the
  detector $A_{b,f}^{model}$ for all the beam energy and filter combinations.
\item Calculate the ratio of measured to predicted $R_{data}^f/R_{model}^f$, 
  normalizing to
  the no-filter case separately for \Eb=2.1\gev\ and 4.0\gev.
\item Perform a two-parameter fit to the 11 normalized double ratios 
  of measured to
  predicted image areas, where the first parameter is the constant value best
  matching the ratios for \Eb=2.1\gev, and the second is the constant 
  value best matching the points for \Eb=4.0\gev. These two values may be
  different from one another, but ideally would be close to unity.
\item Calculate the $\chi^2$ for the 9 degrees of freedom of the fit.
\item Observing which filter points contribute most to $\chi^2$, 
  use \Fig{fig:trans} to determine where the model's detector
  response could be altered to improve the fit.
\end{itemize}

\begin{figure}[t]
   \includegraphics*[width=\figwid]{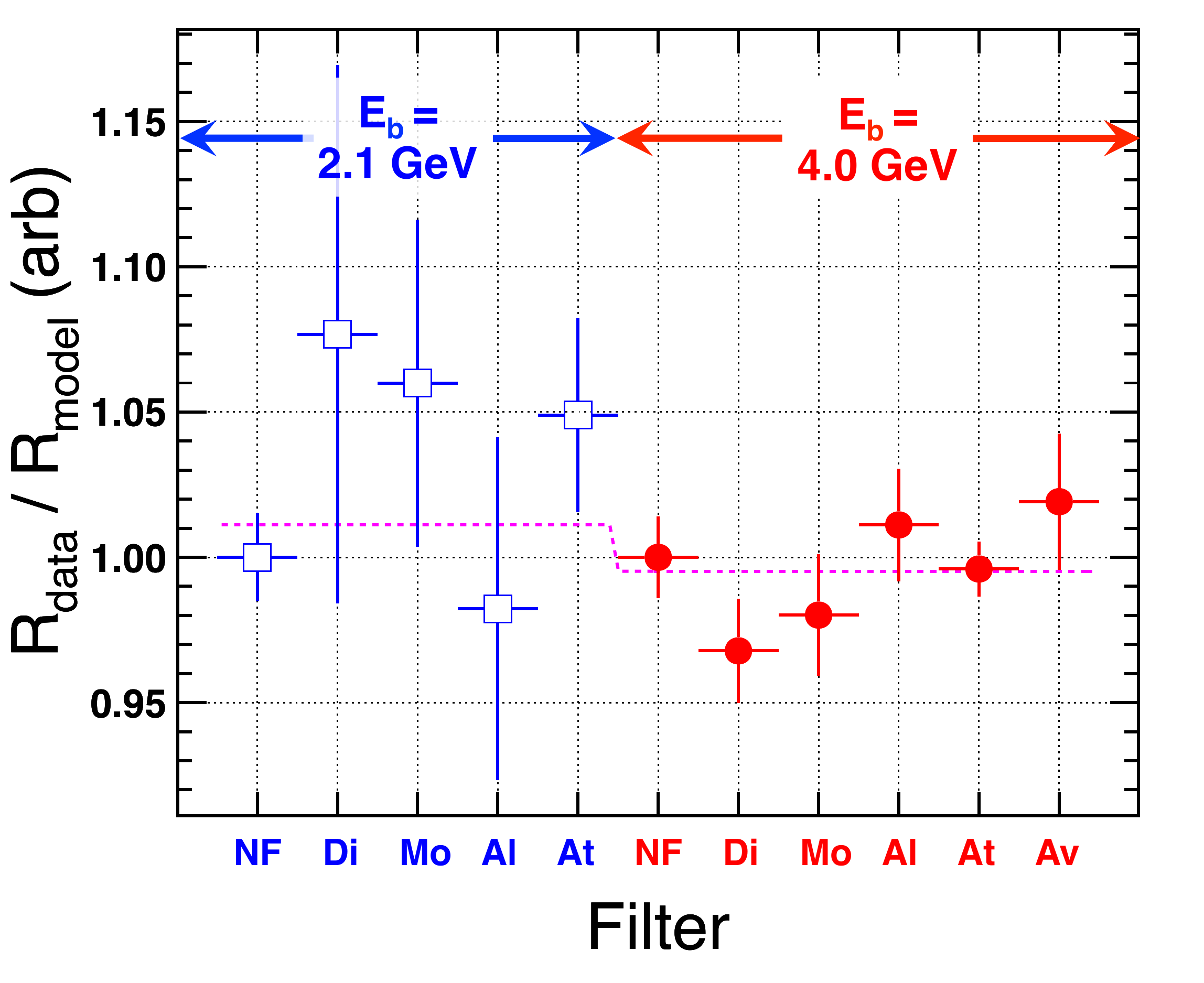}
   \caption{Double ratios $R_{data}^f/R_{model}^f$ of measured to predicted 
     image areas for \Eb=2.1\gev\ (left, open squares) and 4.0\gev\ 
     (right, closed circles), along with the best-fit
     values for each energy range shown by the dashed line.}
\label{fig:tunespectrum}
\end{figure}

\begin{figure}[t]
   \includegraphics*[width=\figwid]{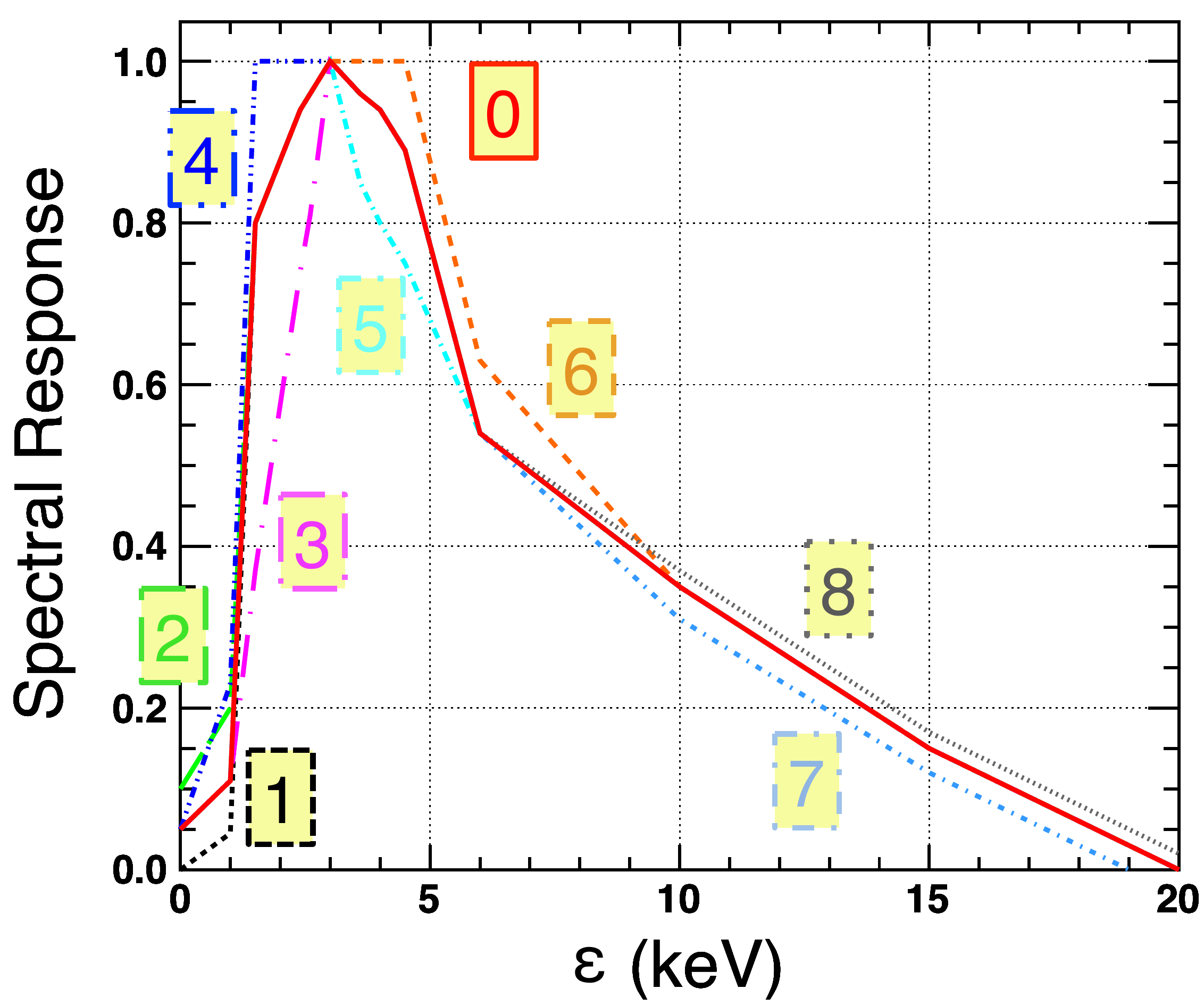}
   \caption{The spectral response of the pixel detector measured from 2.1\gev\
     and 4.0\gev\ data, as described in the text, is shown as the solid line, 
     and labeled by the number 0. Other labeled lines show alternate 
     responses, each of which, by itself, is approximately $\sim 1\sigma$ 
     from nominal. The unseen portions of the lines labeled other than 0
     coincide with the nominal response.}
\label{fig:specenv}
\end{figure}

\begin{figure}[t]
   \includegraphics*[width=\figwid]{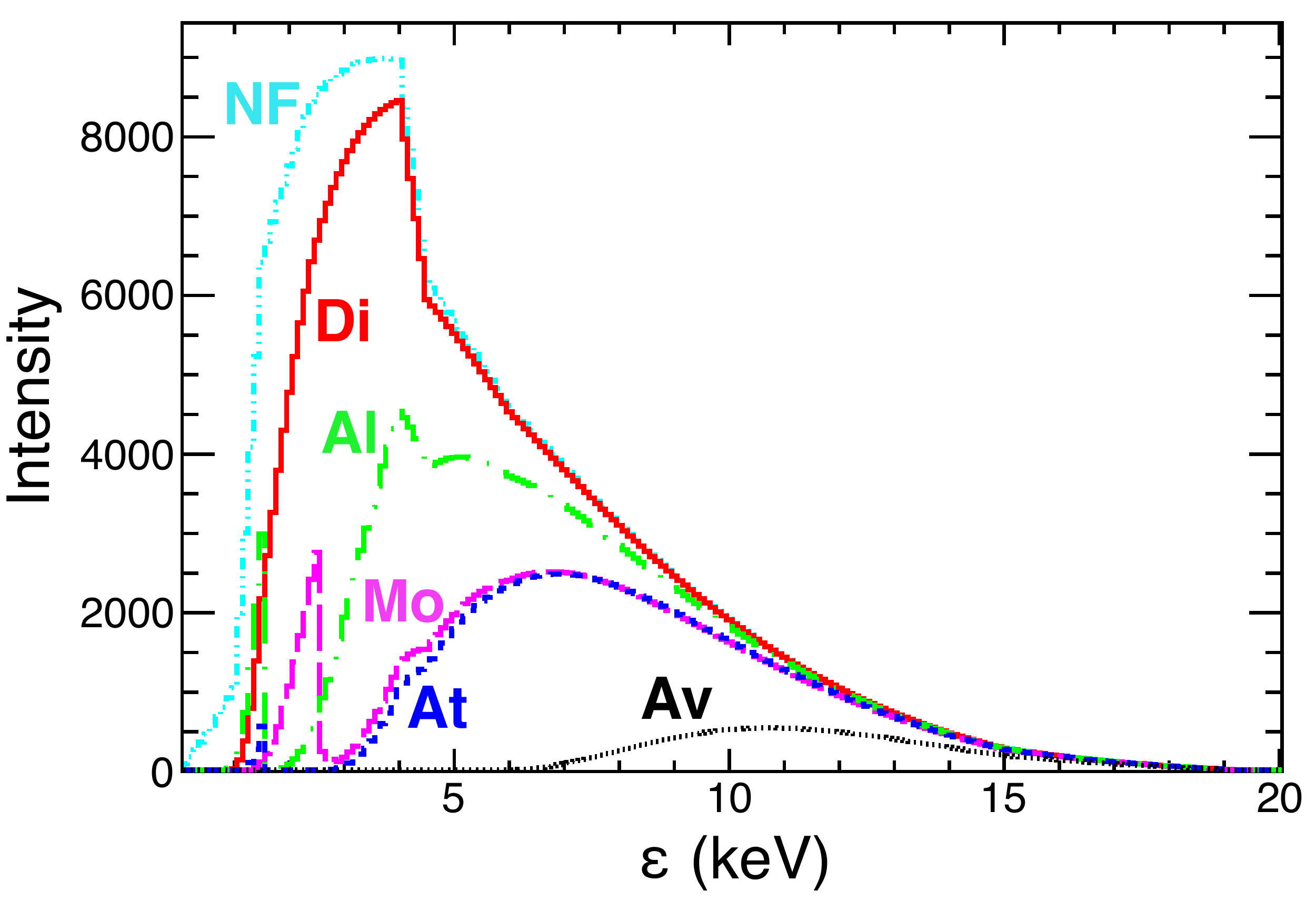}
   \caption{Detected intensity spectrum as a function of x-ray energy for the 
     six filter configurations shown, after accounting for absorption of the
     filters and the inherent spectral response of the detector.}
\label{fig:intensity}
\end{figure}

  The model-normalized data points and fit for the iterative model 
optimization are shown in \Fig{fig:tunespectrum}. The fit has 
$\chi^2=8.0$ for 9 degrees of freedom. The error bars on the 2.1\gev\ points
are dominated by thickness uncertainty due to the filters absorbing a much 
higher relative fraction of the incident energy than at 4.0\gev. 
In order to find a $\sim$1$\sigma$ envelope of possible responses, we
vary the model in limited ranges of x-ray energy until an increase in $\chi^2$ 
of 9 units to $\sim$18 is obtained. The resulting best-fit response is shown 
in \Fig{fig:specenv}, along with alternate responses that are each consistent 
with the data within $\sim$1$\sigma$. \Figure{fig:intensity} shows the
intensity spectrum after accounting for detector spectral response and filter
absorption.

\section[Beam size studies]{Beam size studies}
\label{sec:size}

\begin{figure}[b]
   \includegraphics*[width=\figwid]{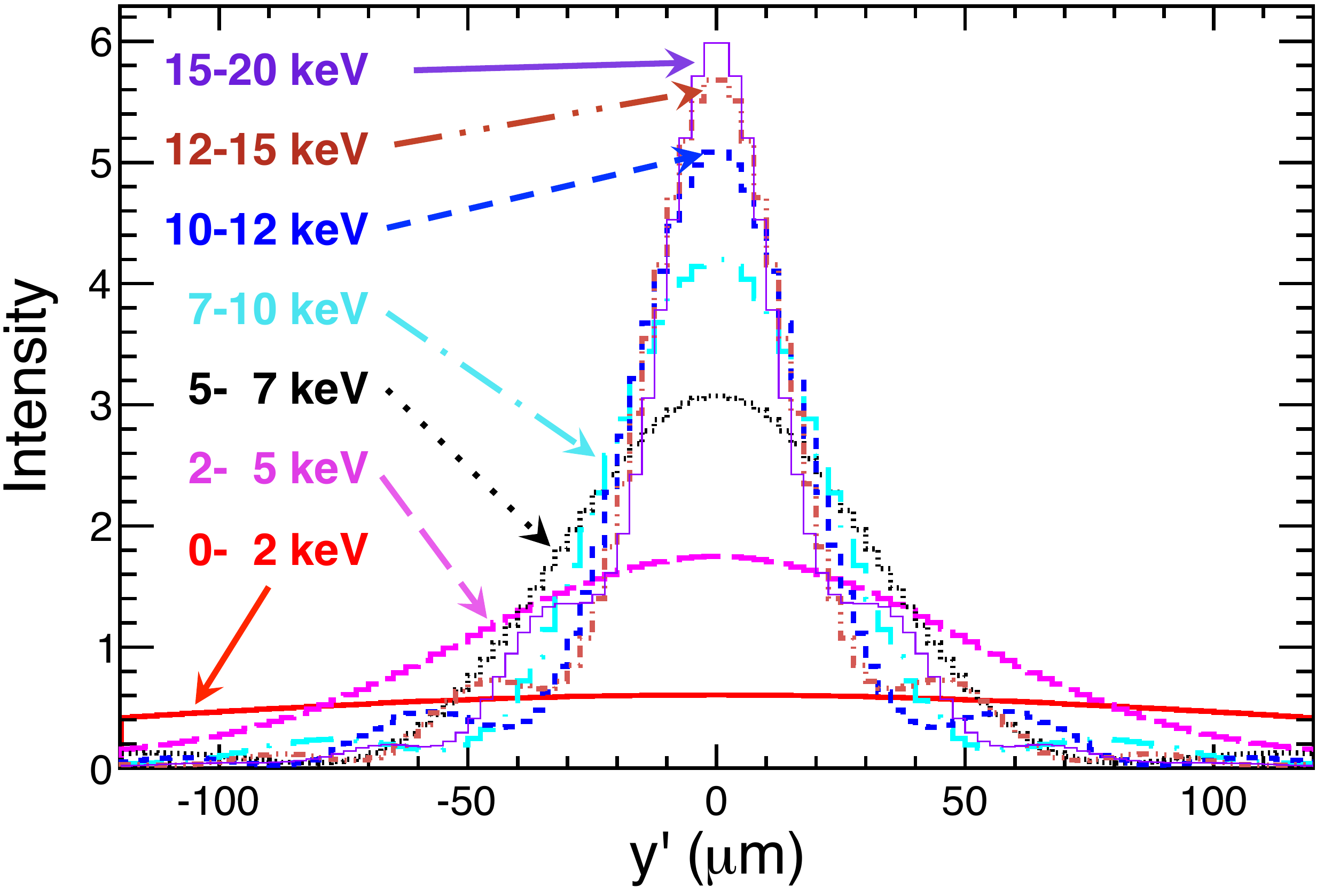}
   \caption{Point response functions (\prftt s) for a pinhole with a 22\mi\ gap
     and no filter in place for the x-ray energy ranges shown. All curves are
     normalized to the same area over the full detector height of 1.6\mm.}
\label{fig:prfs}
\end{figure}

In order to evaluate the effectiveness of the spectral response determined in
\Sec{sec:spec}, we can compare the measured vertical beam with different 
filters in place. The shape of the pinhole \prftt\ varies considerably with
x-ray energy, as seen in \Fig{fig:prfs}, so inaccuracies in the spectral 
response will show up as differences in measured beam size when in fact the beam
size is the same for all filters.

\begin{figure}[t]
   \includegraphics*[width=\figwid]{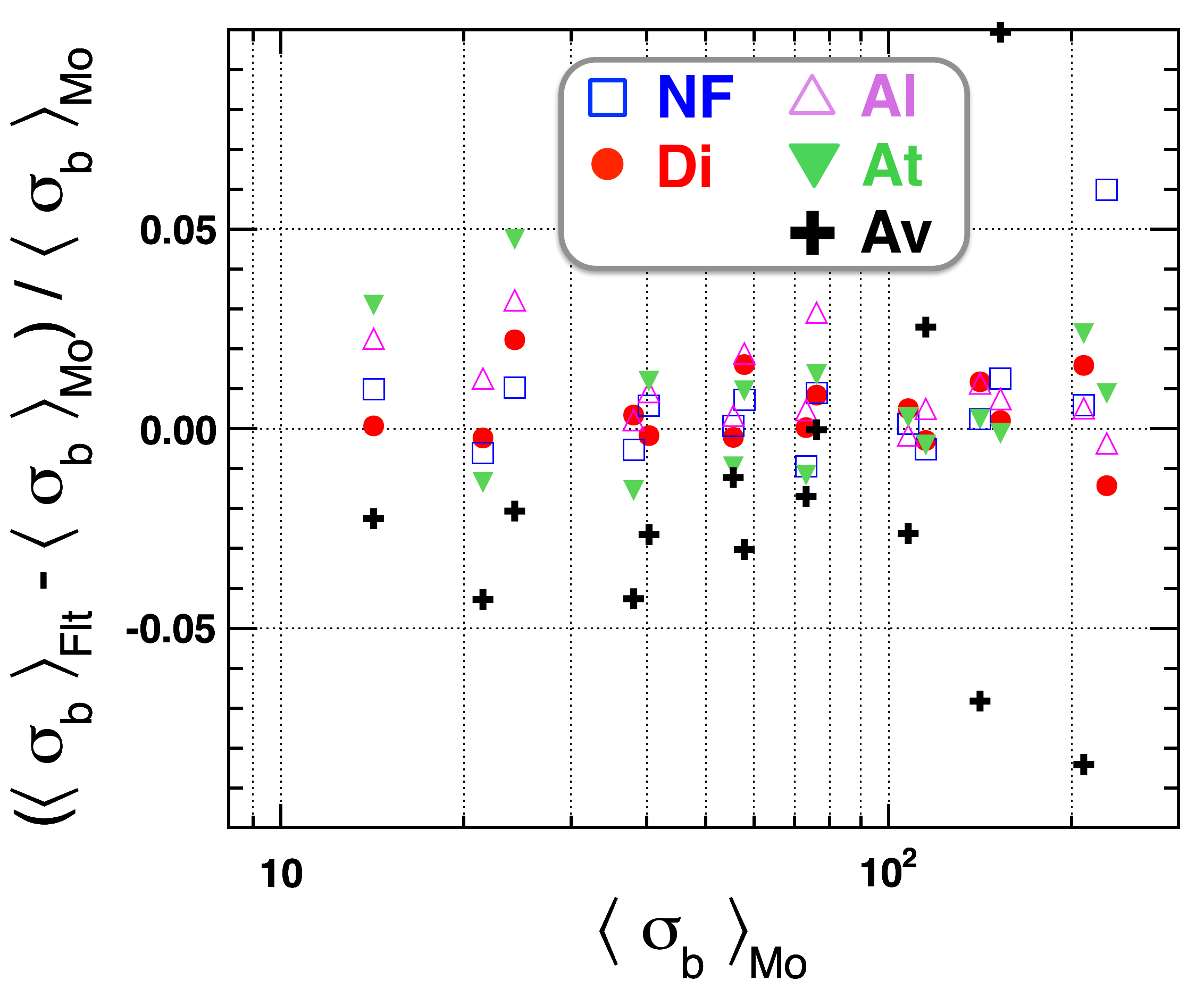}
   \caption{Comparisons of beam sizes obtained with different filters in front
     of the detector as a function of vertical beam size, using the spectral 
     response of \Sec{sec:spec} in creating the \prftt's applicable to each 
     filter. The vertical axis is the difference between the turn-averaged 
     vertical beam size obtained with a given filter minus that obtained at 
     the same emittance setting with the Mo filter, relative to that obtained 
     with Mo. The horizontal axis is the turn-averaged vertical beam size 
     obtained with the Mo filter for each emittance setting.}
\label{fig:sizep}
\end{figure}

The data were acquired with each filter at 15 reproducible settings of CESR beam
emittance (and thus beam size). If the spectral response is correct, beam sizes 
measured with data using different
filters will all yield the same value at the same emittance setting. The
degree of agreement is shown in \Fig{fig:sizep}, where relative differences in
measured beam size are seen to be within several percent of one another,
with a few outliers above a beam size of 120\mi. Accuracy for measuring small
and moderate beam sizes is at the level of $\pm$4\% for all filter
configurations. 

\section[Tilted Detector]{Tilted Detector}
\label{sec:tilt}

The location for a new, dedicated beam size monitor beamline at CESR does not
have enough length available for a conventional design: the magnification 
$M\equiv\lp/L$ (see \Fig{fig:geom}) would be too small, and the detector pitch
would become the limiting factor for measurement of small beam sizes. For such
a short beamline with small magnification, the image from a small beam through 
a pinhole optic would become isolated in just a very few pixels, and resolving 
power at small beam sizes would become poor. To spread out the image over more 
pixels without adding length to the beamline, we have elected to tilt the 
detector forward. A tilt angle of 60$^\circ$ effectively doubles the 
magnification, although the incident light intensity per pixel is halved. 
However, tilting the detector also changes its spectral response. On one hand, a
tilt increases the effective thickness of the SiN passivation layer, cutting 
off light at x-ray energies below $\sim$4\kev; on the other, a tilt increases 
the effective thickness of the active pixel volumes, increasing the 
absorption at x-ray energies above $\sim$4\kev. In order to explore the net 
effect of a tilt, we acquired data at \Eb=2.1 and 4.0\gev\ with the detector 
tilted vertically at 60$^\circ$ and then remeasured the detector spectral 
response.

\begin{figure}[t]
   \includegraphics*[width=\figwid]{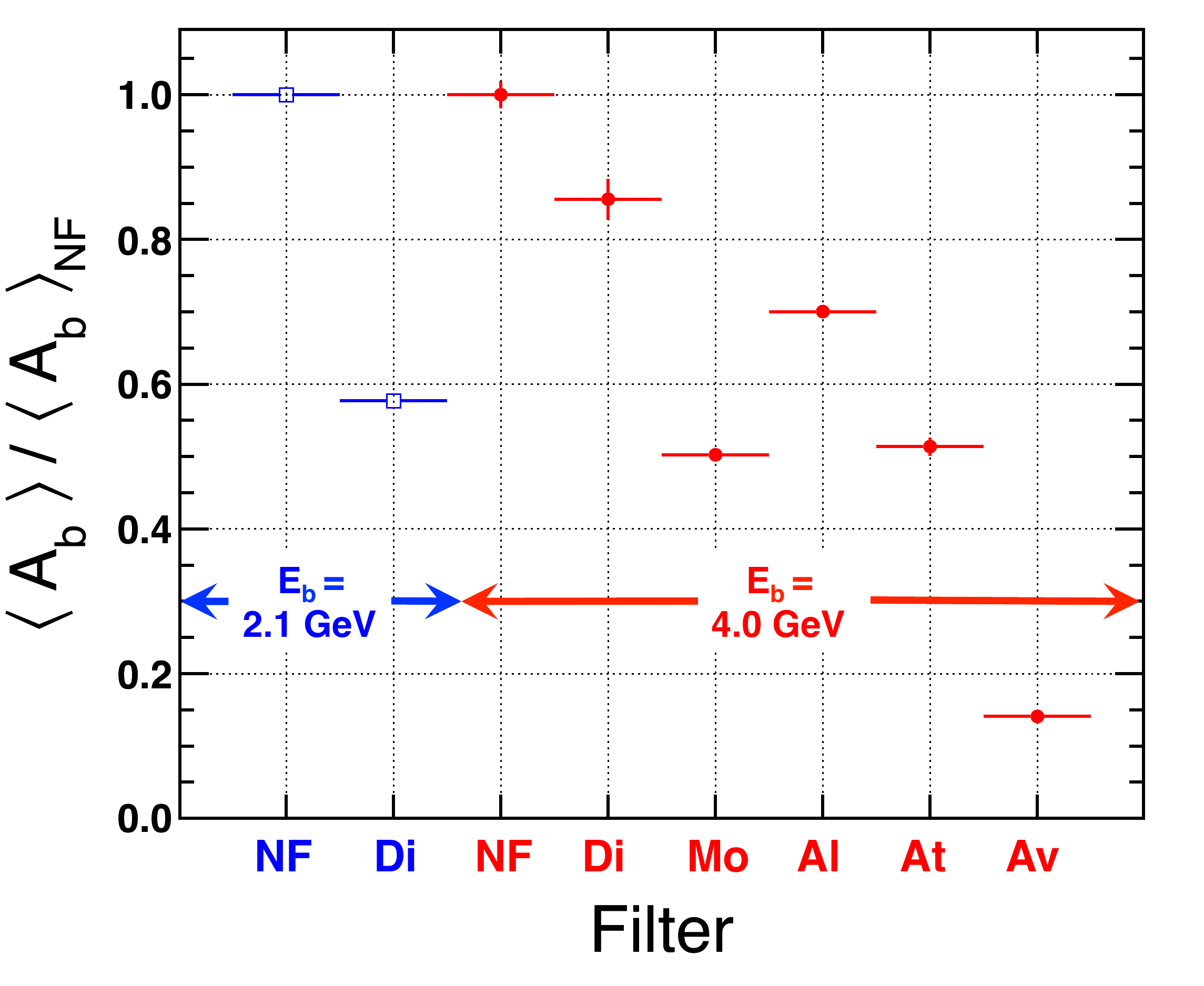}
   \caption{Turn-averaged amplitudes for \Eb=2.1\gev\ (left, open squares) 
     and 4.0\gev\ (right, closed circles) for data acquired with a
     60$^\circ$ detector tilt. Each set has been normalized to its
     own arbitrary scale where the no-filter data point has a value of 1.0.
     Vertical error bars are smaller than the symbols.}
\label{fig:tiltratios}
\end{figure}

\begin{figure}[t]
   \includegraphics*[width=\figwid]{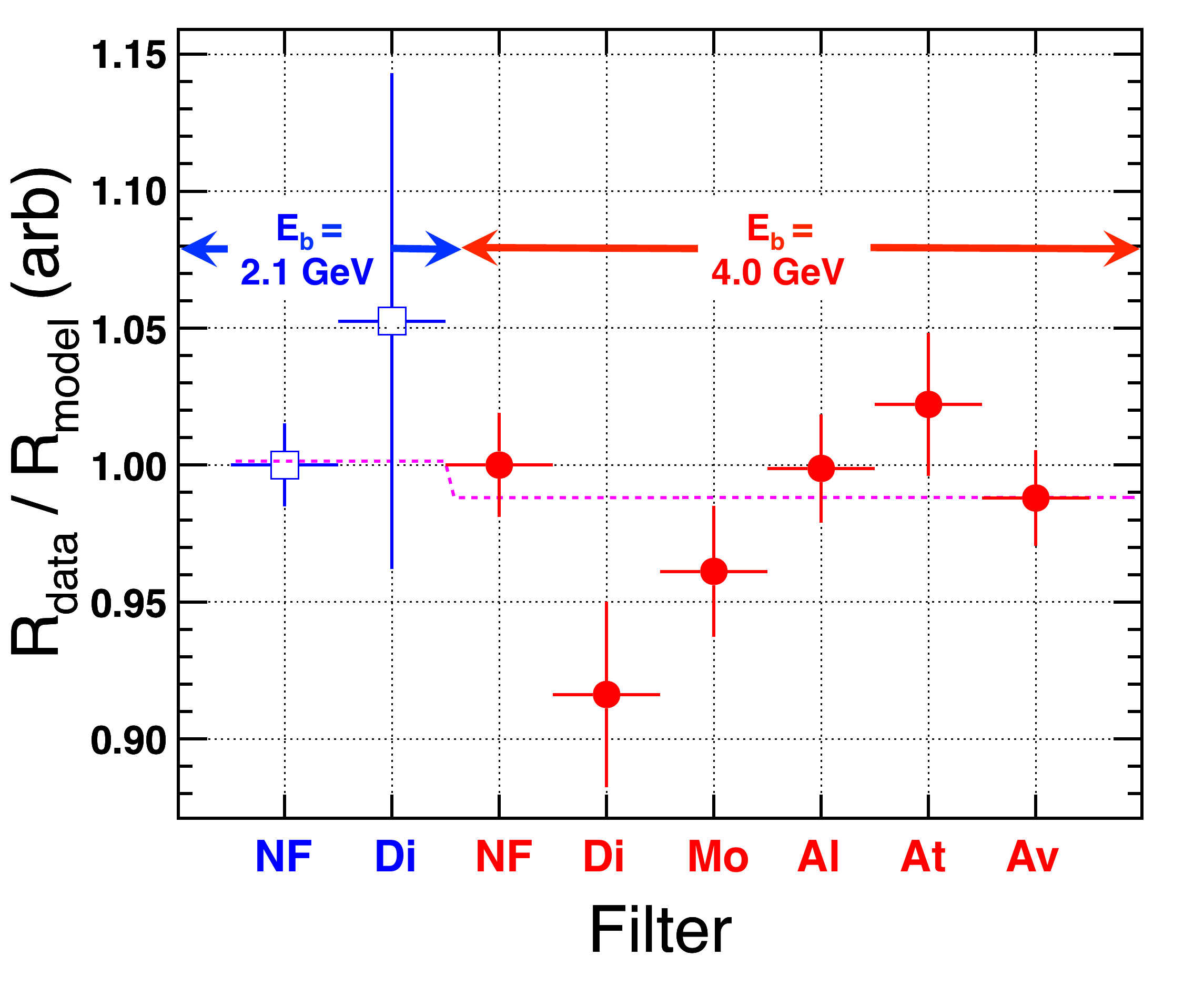}
   \caption{From data acquired with a 60$^\circ$ detector tilt: 
     double ratios $R_{data}^f/R_{model}^f$ of measured to predicted 
     image areas for \Eb=2.1\gev\ (left, open squares) and 4.0\gev\ 
     (right, closed circles), along with the best-fit
     values for each energy range shown by the dashed line.}
\label{fig:tiltFit}
\end{figure}

\begin{figure}[t]
   \includegraphics*[width=\figwid]{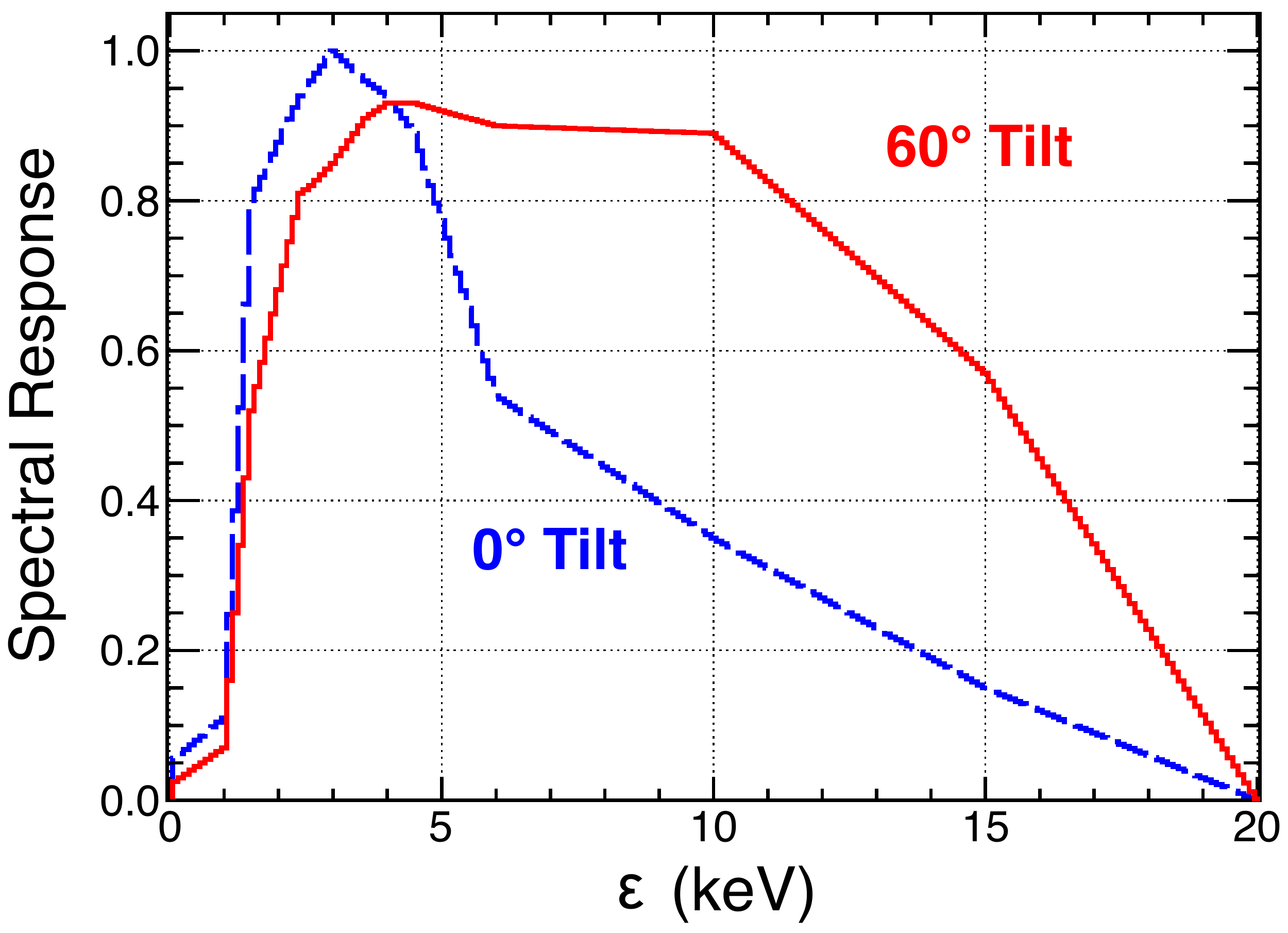}
   \caption{The spectral response of the pixel detector tilted forward by
     60$^\circ$, as measured from 2.1\gev\ and 4.0\gev\ data, is shown as the 
     solid line. The dashed line corresponds to the spectral response 
     determined for no tilt, as shown in \Fig{fig:specenv} as the solid line. 
     Relative normalizations for the two curves is arbitrary and not 
     determined by this measurement; however, we expect the two curves to 
     intersect in the 3-6\kev\ region.}
\label{fig:tiltSpec}
\end{figure}

These data were taken using $e^-$ instead of $e^+$ as in preceding sections. 
Due to limited beam time available, only two filter configurations (NF and Di) 
were implemented for the \Eb=2.1\gev\ data, but the full complement of six 
sets were acquired for the \Eb=4.0\gev\ data. The pinhole opening was set at 
\ap=49\mi, the optimal value for \Eb=2.1\gev, for both datasets. The ratios 
of fitted amplitude per unit current are shown in \Fig{fig:tiltratios}, and 
the resulting fit to the minimum-$\chi^2$ spectral response in 
\Fig{fig:tiltFit}. The minimum-$\chi^2$ spectral response itself appears in 
\Fig{fig:tiltSpec}, where it is compared to that determined for no tilt.

The spectral response for the 60\degree-tilted detector differs qualitatively 
from the no-tilt case in the two expected ways: light absorption is lower (by
about 30\%) below 4\kev\ (due to the thicker passivation layer) and 
higher (by about 100\%) above 4\kev. Thus, although tilting the detector 
spreads incident light over twice the area compared to no-tilt, it recovers 
absorption from x-rays above 4\kev, resulting in a net increase for incident 
spectra that have a substantial above-4\kev\ component.

In order to compare beam size capability, three runs were first taken with a
particular set of CESR parameters at \Eb=2.1\gev\ for the detector in the 
no-tilt configuration. The average value of the beam size was 19.3\mi. 
A short time later, three runs with the 60\degree-tilt were taken for the same
beam conditions; these runs had a mean beam size of 19.0\mi, well within the
typical run-to-run variation of beam size measurements. Hence, we have
demonstrated that spreading out the x-rays by tilting the detector is a viable 
method to increase effective magnification of our pinhole camera as long as a
substantial part of the incident spectrum is above 4\kev.

\section[Coded aperture]{Coded aperture}
\label{sec:cahe}

A coded aperture can be useful as an optical 
element in an x-ray beam-size monitor~\refboth\ by improving the resolving 
power relative to a pinhole. We have found~\reftwo\ that a carefully designed 
coded aperture improves performance only at small beam size because at larger 
beam sizes the multiple peaks in the \prftt\ get merged together, whereas at 
smaller beam size they remain resolved from one another. A coded aperture
allows through more light than a pinhole, and uses a larger fraction of the
detector by spreading several peaks across it. Downsides of coded apertures can
include ineffective patterns for a given incident x-ray energy spectrum, 
inability to withstand heat loads of intense and/or energetic beams, and 
heightened sensitivity (relative to a pinhole) to the alignment of the beam 
with the coded aperture's center.

\begin{figure}[t]
   \includegraphics*[width=\figwid]{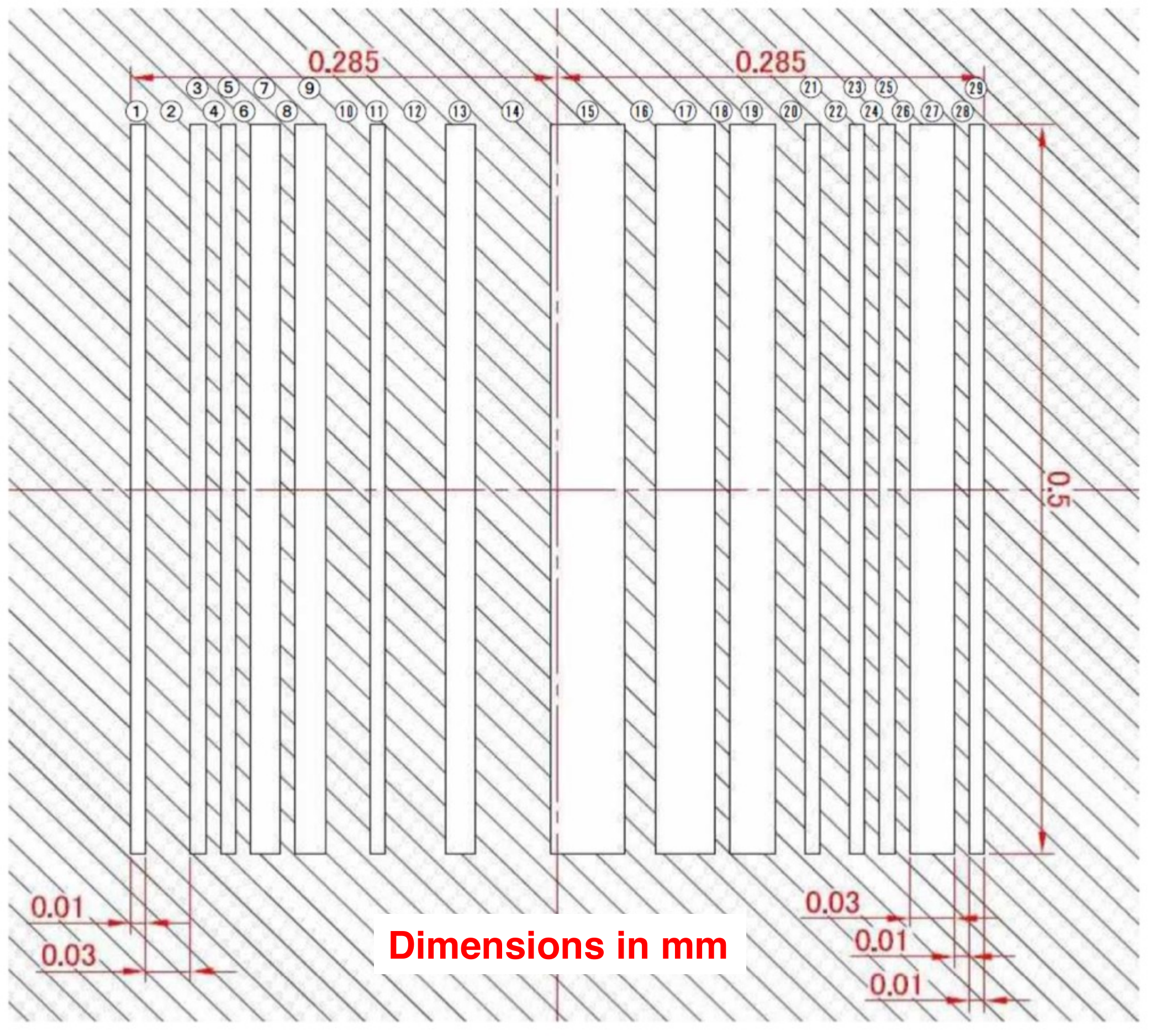}
   \caption{Dimensions of the pattern of the \Eb=4.0\gev\ coded aperture
     CAHE, which consists of a 350\mi-thick diamond backing with an
     8\mi-thick gold (plus 0.1\mi\ chromium) masking (the masking represented 
     by the hatched area).}
\label{fig:pattern}
\end{figure}

Here we report briefly on exposure of a coded aperture designated CAHE. The
dimensional specifications for the CAHE pattern appear in \Fig{fig:pattern}. 
The backing is synthetic diamond 350\mi\ thick; the masking is 8\mi\ of gold 
and 0.1\mi\ of chromium; the manufacturer is Cornes Technology~\cite{ref:cvd}. 
The CAHE optical element is a prototype being evaluated for use in the x-ray 
beam size monitor for SuperKEKB.

\begin{figure}[t]
   \includegraphics*[width=\figwid]{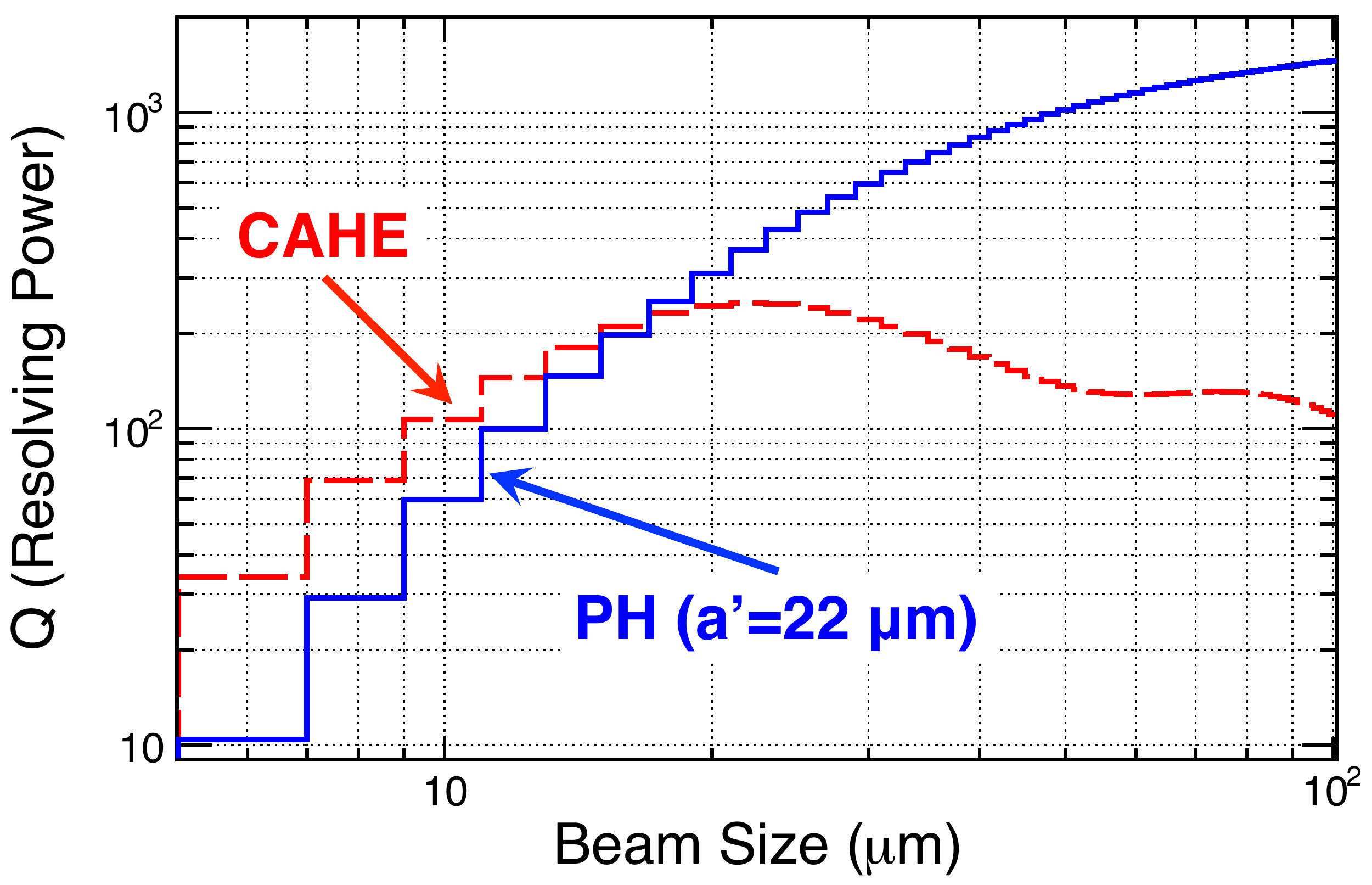}
   \caption{The predicted resolving power $Q$~\refone\ as a function of 
     beam size at \Eb=4.0\gev\ for a pinhole (PH) with \ap=22\mi\ (solid line) 
     and the CAHE coded aperture (dashed line).}
\label{fig:q}
\end{figure}

\begin{figure}[t]
   \includegraphics*[width=\figwid]{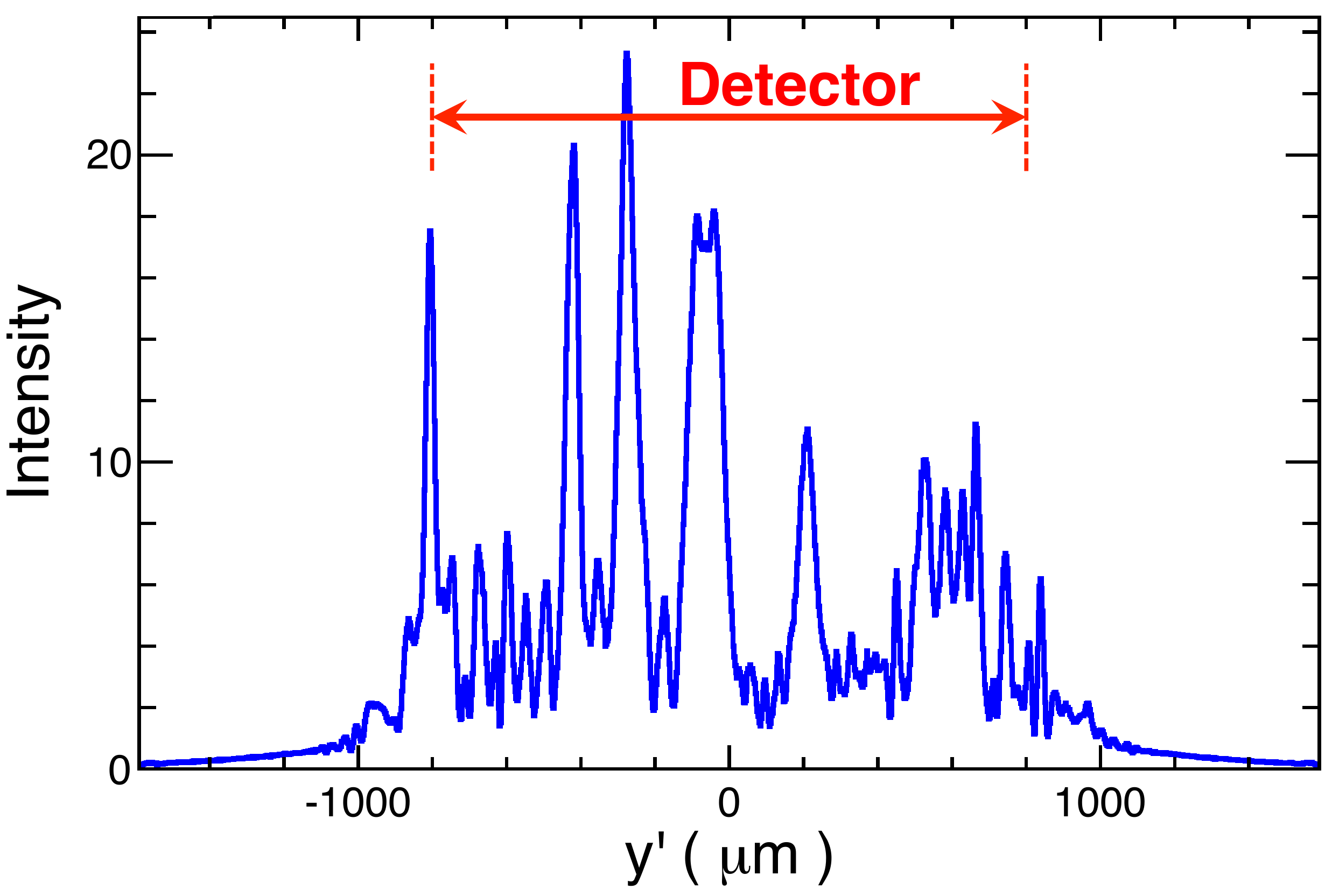}
   \caption{Point response function (\prftt) for the CAHE coded aperture in the
     CESR positron beam at 4.0\gev. The \prftt\ overflows our detector; the
     marked region represents the approximately central exposure detailed in 
     the text.}
\label{fig:prfcahe}
\end{figure}

\begin{figure}[t]
   \includegraphics*[width=\figwid]{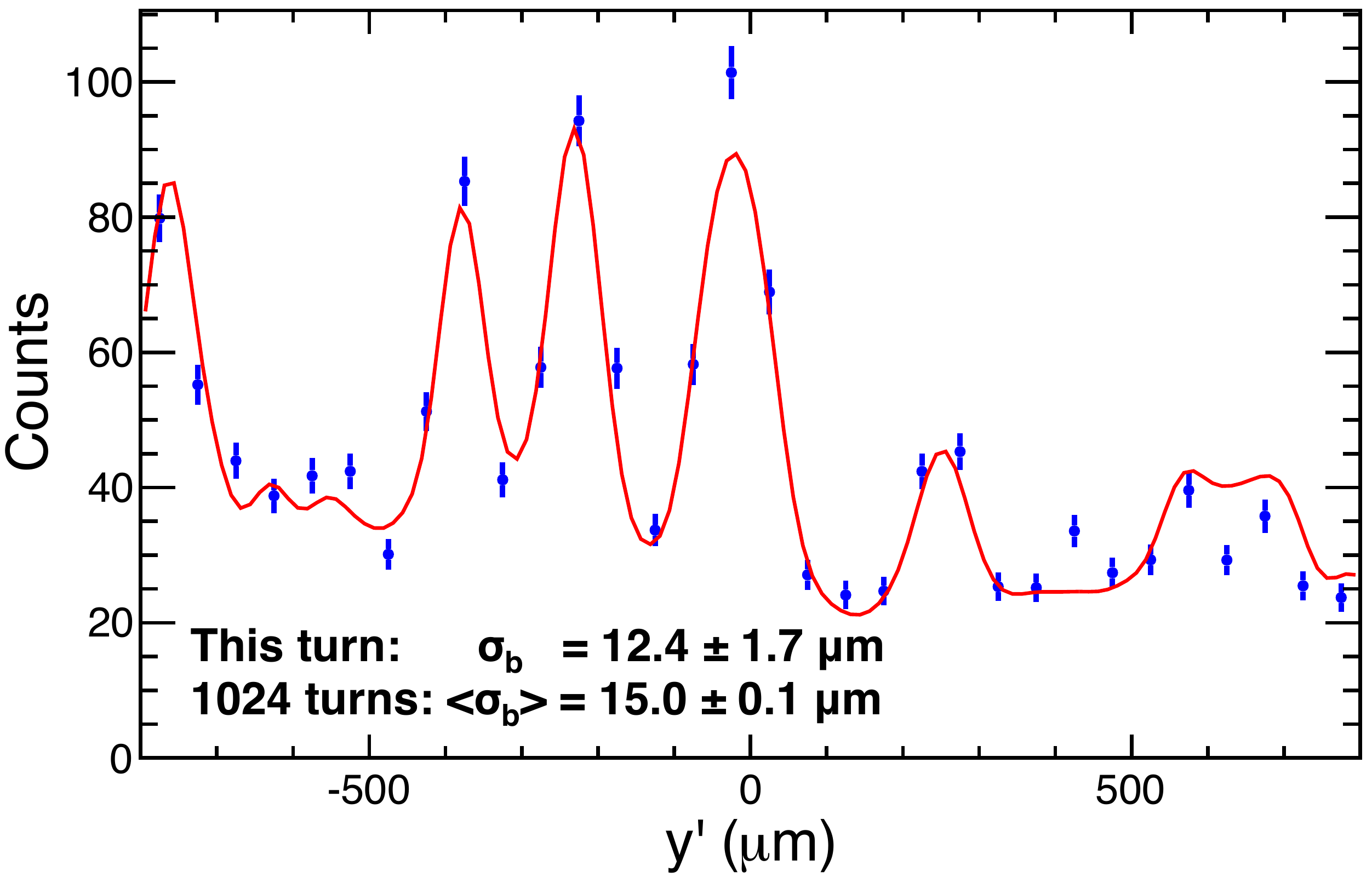}
   \caption{CAHE image from one turn at \Eb=4.0\gev\ overlaid with a fit to 
     beam size, position, and amplitude. The fitted beam size for this turn 
     and the size averaged over all turns in this run are both shown.}
\label{fig:imagecahe}
\end{figure}

\Fig{fig:q} shows the predicted resolving power of CAHE and a 22\mi\ pinhole 
as used in the CESR positron beam at \Eb=4.0\gev. CAHE is expected to yield
better performance than this pinhole below about \siz=15\mi. Unfortunately, the
smallest vertical beam size attained at CESR for 4.0\gev\ positrons is about
\siz=15\mi, so we can directly compare the pinhole and CAHE only at and above 
the beam size where the model predicts that they are equivalent in resolving 
power. The \prftt\ of CAHE is shown in \Fig{fig:prfcahe}.

Data were acquired at 4.0\gev\ with a positron beam on CAHE for the same
emittance settings as for the pinhole described previously. We found that beam
sizes above about 25\mi\ were not reproduced well by CAHE and so focus only on
the smallest beam size attained, \siz$\approx$15\mi. The CAHE images were fit 
using a fixed detector offset of $d$=1\mm\ (implying that the beam was 
vertically offset from the center of the CAHE optic by $d/M=417$\mi). 
\Figure{fig:imagecahe} shows an image of a particular turn at the same beam 
conditions where the pinhole-determined beam size, averaged over filters, was 
about 14.3\mi. The turn shown yields a beam size of \siz=12.4$\pm$1.7\mi, and 
the 1024-turn-averaged beam size is \sizta=15.0$\pm$0.1\mi, larger than
the pinhole average by $\approx$5\%.

\section[Conclusions]{Conclusions}
\label{sec:conc}

We have extended the characterization of the InGaAs strip detector in the 
CESR-TA vertical beam size monitor up to an x-ray energy of $\sim$20\kev. The
spectral response so determined allows determination of the vertical beam size
to within $\pm$4\% for data acquired using the pinhole optical element at 
\Eb=4.0\gev. A diamond-backed coded aperture (a prototype optic for 
beam size monitors at SuperKEKB), 
was tested for measuring beam size accurately and found to yield
accurate ($<$5\%) results only for small beam sizes (\siz$<$25\mi). 
Measurements taken for a detector pitched forward by 60\degree\ showed that 
we can double the effective magnification of a new, dedicated xBSM beamline 
for CESR without physically extending the beamline itself.

\section*{Acknowledgments}

{\small
This work would not have been possible without the dedicated and skilled 
efforts of
the CESR Operations Group as well as the support of the Cornell Laboratory for
Accelerator-based Sciences and Education (CLASSE) and Cornell High Energy 
Synchrotron Source (CHESS). This research was supported under the National 
Science Foundation awards PHY-1002467 and PHY-1416318, and by the NSF and 
National Institutes of Health/National Institute of General Medical Sciences 
under NSF award DMR-1332208. It was also supported by the
Japan-US Cooperation Program in High Energy Physics.
}

%\clearpage

\end{document}